\documentclass[structabstract]{aa}  
\usepackage{graphicx}
\usepackage{txfonts}
\usepackage{natbib}
\begin{document}

\title{Small-scale systems of galaxies. IV. } 

\subtitle{Searching for the faint galaxy population \\
associated with X-ray detected isolated E+S pairs}

\author{R.~Gr\"utzbauch \inst{1,2} 
   \and W.W.~Zeilinger \inst{1} 
   \and R.~Rampazzo \inst{3} 
   \and E.V.~Held \inst{3}
   \and J.W.~Sulentic \inst{4}
   \and G.~Trinchieri \inst{5}}

\institute{Institut f\"ur Astronomie, Universit\"at Wien,
           T\"urkenschanzstra{\ss}e 17, A-1180 Wien, Austria 
      \and University of Nottingham, School of Physics and Astronomy, Nottingham NG7 2RD, UK 
      \and INAF - Osservatorio Astronomico di Padova, Vicolo
           dell'Osservatorio 5,I-35122, Padova, Italy
      \and Department of Physics and Astronomy, University of Alabama,
           Tuscaloosa, AL 35487, USA
      \and INAF - Osservatorio Astronomico di Brera, Via Brera 28,
           I-20121, Milano, Italy}

\abstract
{In hierarchical evolutionary scenarios, isolated, physical pairs may represent an intermediate phase, or ``way station'', between collapsing groups and isolated elliptical (E) galaxies (or fossil groups).} 
{We started a comprehensive study of a sample of galaxy pairs composed of a giant E and a spiral (S) with the aim of investigating their formation/evolutionary history from observed optical and X-ray properties.}
{We present VLT-VIMOS observations designed to identify faint galaxies associated with the E+S systems from candidate lists generated using photometric criteria on WFI images covering an area of $\sim$ 0.2 $h_{100}^{-1}$ Mpc radius around the pairs.}
{We found two and ten new members likely to be associated with the X-ray bright systems RR~143 and RR~242, respectively. The X-ray faint RR~210 and RR~216, which were only partially covered by the VIMOS observations, have two and three new faint members, respectively. The new members increase the number of associated galaxies to 4, 7, 6, and 16 for RR~143, RR~210, RR~216, and RR~242, respectively, down to $M_R \sim -12 + 5 \log h_{100}$. We derive structural properties of the faint members from surface photometry. The faint galaxy population of all the systems is dominated by disk galaxies, 40\% being S0s with generally low bulge to total light ratios. A small fraction of the faint companions show signatures of interaction. A remarkable shell system is detected in the early-type galaxy RR~242\_24532. We also derive dynamical properties and optical luminosity functions for the 4 groups.}
{The above results are discussed in the context of the evolution of poor galaxy group associations. A comparison between the Optical Luminosity Functions (OLFs) of our E+S systems and a sample of X-ray bright poor groups suggest that the OLF of X-ray detected poor galaxy systems is not universal. The OLF of our X-ray bright systems suggests that they are more dynamically evolved than our X-ray faint sample and some X-ray bright groups in the literature. However, we suggest that the X-ray faint E+S pairs represent a phase in the dynamical evolution of some X-ray bright poor galaxy groups. The recent or ongoing interaction in which the E member of the X-ray faint pairs is involved could have decreased the luminosity of any surrounding X-ray emitting gas.}
{}


\maketitle

\section{Introduction}

The X--ray signature of a hot Intra-Group Medium (IGM) has been detected in loose groups containing an early-type galaxy population \citep[e.g.,][]{M00}.  The pioneering work of \citet{Z99} suggested that groups might fall into different classes defined by their X--ray properties: from groups with a luminous, extended, hot IGM centred on a giant E to groups with little or no diffuse emission. Several examples of these classes can now be found in the literature \citep[see e.g.,][] {Mul03,Tri03,Bel03,Ota04}. In a hierarchical evolutionary scenario, the final product of a merged group would be a luminous isolated E with an extended X-ray halo, and a few have indeed been identified \citep[see e.g.,][]{Mul99,Vik99, Jon03,Kho04}. \citet{Jon03} estimated the incidence of fossil groups. They found that fossil systems, defined as a spatially extended X-ray source with an X-ray luminosity from diffuse, hot gas of L$_{X,bol} \geq 10^{42}$ h$_{50}^{-2}$ erg s$^{-1}$, represent 8-20\% of all systems of the same X-ray luminosity.
However, an optical study of a sample of 100+ isolated early-type galaxies found that almost no systems were luminous enough to have been the product of a merger between galaxies brighter than $L^\ast$, i.e., a merged group \citep{Sulentic06}.
Chandra and XMM-Newton observations of optically selected merger remnants show that their hot gas is X-ray underluminous compared with mature E galaxies into which these merger remnants are expected to evolve \citep[see e.g.,][]{Sansom00,Nol04,Rampazzo06}. \citet{Brassington07},  investigating the evolution of X-ray emission during the merger process, similarly found that just after an accretion episode ($\sim$1 Gyr after coalescence) merger remnants are X-ray faint compared to a typical mature E galaxy. They suggested that these systems will start to resemble typical elliptical galaxies at a greater dynamical age (after $\sim$3 Gyr), supporting the idea that halo regeneration takes place within low L$_X$ merger remnants.  

Compact galaxy groups generally show modest diffuse X-ray emission \citep[e.g.,][]{Trinchieri05}. However, optically selected structures (such as compact groups) generally tend to be X-ray underluminous in comparison to X-ray selected systems \citep{Pop07,Ras06} and consequently the modest diffuse X-ray emission is not necessarily associated with a recent galaxy merger. \citet{Ras06} found that low level IGM emission could be an indication that the group is in the process of collapsing for the first time. Other possibilities include either that the gravitational potentials are too shallow for the gas to emit substantially in X-rays or that there is simply little of no intra-group gas present in those groups.    

We have extended the optical and X-ray studies to isolated physical pairs of galaxies which are simple, and rather common galaxy aggregates in low density environments (LDEs). Among pairs, the mixed E+S binary systems are particularly interesting in the context of an evolutionary accretion scenario
\citep[see e.g.,][]{RS92,HT01,Dom2003,Dom05}, since the luminous E components might be merger products.  The study of such relatively simple structures may then shed light on a {\it possible evolutionary link} between poor groups and isolated Es.  One of the most spectacular examples involves the optical
and X-ray bright isolated E+S pair (CPG127=Arp114=NGC2276+2300) with only two known dwarf companions \citep{Davis96}.

In this context, we initiated an optical and X-ray study of four E+S systems: RR~143, RR~210, RR~216, and RR~242 \citep[][hereafter Paper~III] {Gru07}. In contrast to their similar optical characteristics, their X-ray properties \citep[see also][]{TR01} indicate that their X-ray luminosities, L$_X$/L$_B$ ratios and morphologies are very different, which implies that they have different origins and/or represent different evolutionary stages of the systems.
X-ray emission in RR~143 and RR~242 is centred on the elliptical but is much more extended than the optical light.
The emission in RR~143, although centred on the elliptical, shows an asymmetric elongation towards the late-type companion. The total extension is
r $\sim$ 500\arcsec\ (90 $h_{100}^{-1}$~kpc).  The extended emission from RR~242 is more regular and has an extent  of 700\arcsec\ (120 $h_{100}^{-1}$~kpc). RR~210 and RR~216 show relatively faint and compact (i.e., within the optical galaxy) X-ray emission, consistent with an origin in an evolved stellar population.  The emission in RR~143 and RR~242 can be argued to be related to a group potential (as in CPG127) rather than to an individual galaxy. 
In such a scenario, RR~210 and RR~216 could represent the {\it active} part of very poor and loose {\it evolving} groups \citep[see e.g.,][]{Rampazzo06}. 
The {\it activity} is reflected by the optically extended and distorted morphologies (see Paper~III).

This paper presents results of VLT-VIMOS observations of the faint galaxy populations around the above four RR E+S pairs. Candidates were selected based on their magnitude, $(V-R)$ colour, and size in Paper~III.  
The new observations allow us to determine the redshifts of these faint galaxies and consequently their membership in the E+S systems. These measurements enable us to discuss these groups in the context of previous work by \citet{ZM98}. They found a significantly higher number of faint galaxies ($\sim$ 20 - 50 members with $M_B \leq -14 +5 \log h_{100}$) in groups with a significant hot IGM compared to groups without this component. We estimate optical luminosity functions (OLF hereafter) for the combined X-ray bright and X-ray faint groups, respectively, and evaluate them in the context of group evolution \citep[see e.g.,][]{ZM00}.  We also compare the characteristics of the galaxy population in the E+S pairs' environments with those of other X-ray detected groups \citep[see e.g.,][]{Tran01}.

The paper is organised as follows. Section~\ref{obsandred} describes the VLT-VIMOS observations as well as data reduction techniques.  Results are presented in section~\ref{results}. A discussion of the results in the light of current literature is given in section~\ref{discussion}.

\begin{table*}
\centering
\begin{tiny}
\caption{Log of VIMOS observations}
\begin{tabular}{llcccccc}
\hline\hline
Field & OB ID & Observation date & Exposure time & Airmass & Seeing & Moon dist.$^{1}$ & Lunar illum.$^{2}$\\
& & JJJJ-MM-DD & [sec] & & [$^{\arcsec}$] & [$^{\circ}$] & \\
\hline
RR~143-a & 218955 & 2006-01-29 & 1800 & 1.298 & 0.64 & 90.7 & 0.005 \\
RR~143-a & 218955 & 2006-01-29 & 1800 & 1.308 & 0.65 & 90.8 & 0.004 \\
RR~143-b & 218973 & 2006-01-23 & 1800 & 1.297 & 0.84 & 85.2 & 0.446 \\
RR~143-b & 218973 & 2006-01-23 & 1800 & 1.306 & 0.78 & 85.2 & 0.444 \\
RR~143-c & 218964 & 2006-01-30 & 1800 & 1.359 & 0.85 & 92.1 & 0.006 \\
RR~143-c & 218964 & 2006-01-30 & 1800 & 1.408 & 0.69 & 92.1 & 0.006 \\
RR~143-d & 218946 & 2006-01-22 & 1800 & 1.319 & 0.83 & 85.1 & 0.550 \\
RR~143-d & 218946 & 2006-01-22 & 1800 & 1.304 & 0.82 & 85.1 & 0.547 \\
& & & & & & & \\
RR~210-a & 219249 & 2006-03-31 & 1800 & 1.018 & 0.66 & 150.3 & 0.041 \\
RR~210-a & 219249 & 2006-03-31 & 1800 & 1.006 & 0.66 & 150.1 & 0.042 \\
RR~210-b & 219258 & 2006-03-26 & 1800 & 1.073 & 1.03 & 121.6 & 0.138 \\
RR~210-b & 219258 & 2006-03-26 & 1800 & 1.135 & 1.05 & 121.9 & 0.136 \\
& & & & & & & \\
RR~216-a & 219285 & 2006-03-29 & 1800 & 1.052 & 1.03 & 143.1 & 0.000 \\
RR~216-a & 219285 & 2006-03-29 & 1800 & 1.081 & 1.03 & 143.3 & 0.000 \\
& & & & & & & \\
RR~242-a & 219759 & 2006-03-24 & 1800 & 1.104 & 0.82 & 76.3 & 0.328 \\
RR~242-a & 219759 & 2006-03-24 & 1800 & 1.150 & 0.94 & 76.6 & 0.326 \\
RR~242-b & 219750 & 2006-03-09 & 1800 & 1.093 & 0.82 & 110.0 & 0.730 \\
RR~242-b & 219750 & 2006-03-09 & 1800 & 1.175 & 0.73 & 109.5 & 0.734 \\
RR~242-c & 219741 & 2006-03-25 & 1800 & 1.208 & 1.40 & 89.2 & 0.221 \\
RR~242-c & 219741 & 2006-03-26 & 1800 & 1.069 & 1.17 & 100.5 & 0.141 \\
RR~242-d & 219732 & 2006-03-26 & 1800 & 1.132 & 1.06 & 101.9 & 0.132 \\
RR~242-d & 219732 & 2006-03-26 & 1800 & 1.191 & 1.09 & 102.1 & 0.131 \\
\hline
\multicolumn{8}{l}{}\\
\multicolumn{8}{l}{$^{1}$ Angular distance of the moon on the sky.} \\
\multicolumn{8}{l}{$^{2}$ Fractional illumination of the moon.} \\
\end{tabular}
\end{tiny}
\label{table1}
\end{table*}


\section{Observations and reduction}\label{obsandred}

The colour selection applied to WIde Field Imager (WFI) images described in Paper~III permitted us to isolate a sample of faint galaxies possibly associated with the E+S systems. 
This sample is referred to as the candidate sample in the following. We summarize briefly the selection criteria used to construct this sample.
Galaxy colours were obtained with {\tt SExtractor} \citep{Bert96}. The source extraction was completed for the R-band and V-band images simultaneously, using the same extraction criteria for both bands. The {\tt MAG\_AUTO} output magnitudes from {\tt SExtractor} were then calibrated using the photometric equations given in Paper~III (Section~4.2). The colour selection was based on the colour-magnitude relation of the Virgo Cluster \citep{Vis77}, from which the expected location of the red sequence at the pairs' distance was computed. The bright member galaxies in the four groups indeed follow this red sequence or are a bit bluer (see Fig.~12 in Paper~III).
\citet{Fuk95} found from synthetic galaxy colours that the K-correction for a typical elliptical galaxy at $z \sim 0.2$ corresponds to a shift in colour of $\Delta (V-R) \sim 0.2$ mag. Adopting an intrinsic colour of bright ellipticals of $(V-R) \sim 0.7$, galaxies with a colour of $(V-R) > 0.9$ are already most likely to be in the background. Therefore, for the sake of simplicity, all objects with $(V-R) \geq 1$ were excluded from the candidate sample. We further applied a general magnitude limit of $m_R = 21$, finter than which the star-galaxy classifier of {\tt SExtractor} becomes unreliable, and a size cut-off at a detected semi-major axis of $a = 1.5^{\prime\prime}$ (see below). This corresponds to $M_R \sim -12 + 5 \log h_{100}$ and $a \sim 400$ pc at the distance of the farthest pair (RR~242).

The candidates are found all over the WFI images ($34^\prime \times 34^\prime$), although in some fields they show peculiar, e.g., clumpy,
distributions (see Fig.~14 in Paper~III).

To cover the entire WFI field of view and obtain spectra with a signal-to-noise ratio adequate for measuring the redshifts of our faint magnitude
candidates, we used VIMOS (VIsible Multi-Object Spectrograph) \citep{LeFevre03} at the Very Large Telescope (VLT) of ESO located at Cerro
Paranal, Chile.  The instrument is mounted on the Nasmyth focus B of UT3 Melipal and has four identical arms, which correspond to the 4 quadrants covering the entire field, each having a field of view of $7^\prime \times 8^\prime$. The gap between each quadrant is $\sim 2^\prime$.

Spectroscopic observations were tailored to derive both the candidate-galaxy redshift and, in a subsequent study, the absorption line--strength indices of member galaxies to investigate their average age and metallicity and infer their star-formation history \citep[see e.g.,][]{Gru05}.  We adopted the HR (high resolution) blue grism, which permits the coverage of the spectral region containing the H$\beta$, Mg2, and Fe ($\lambda$ 5270 \AA, $\lambda$ 5335 \AA) absorption lines with a resolution of R=2050 ($1^{\prime\prime}$ slit) and a dispersion of $0.51$ \AA\ pixel$^{-1}$. Spectrophotometric and Lick standard-stars were either observed or extracted from the VIMOS data archive with the same instrument set-up.  This configuration allows only one slit in the dispersion direction, i.e., each single spectrum covers the full length of the detector.  The wavelength interval depends on the slit position.  At the CCD centre, the wavelength interval is 4150 -- 6200 \AA.  At the upper CCD edge ($+ 4^\prime$), the interval is 4800 -- 6900 \AA\, and 3650 -- 5650 \AA\ at the lower edge ($-4^\prime$).

\begin{figure*}
\includegraphics[width=9.2truecm]{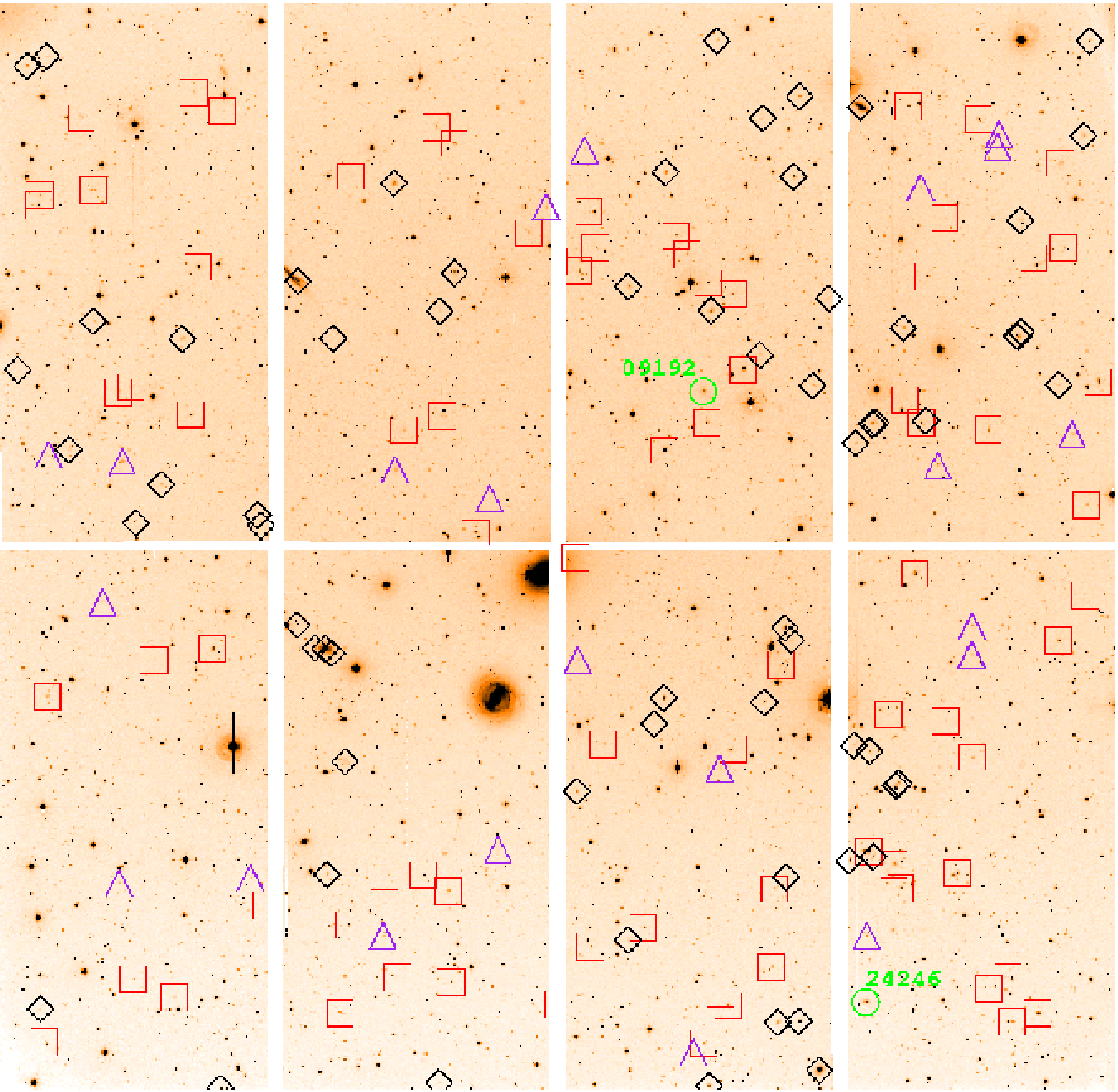}
\includegraphics[width=9.2truecm]{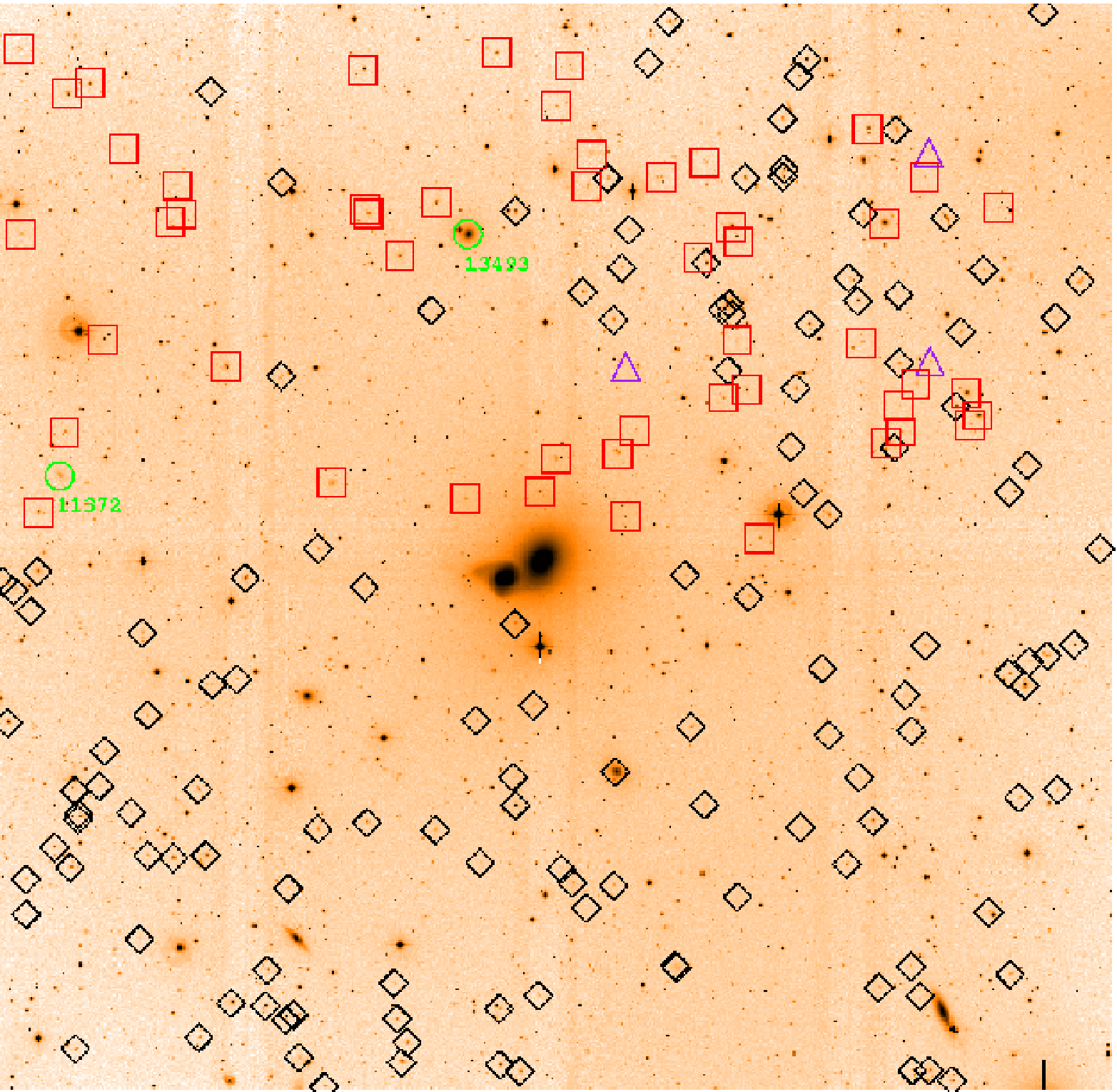}
\includegraphics[width=9.2truecm]{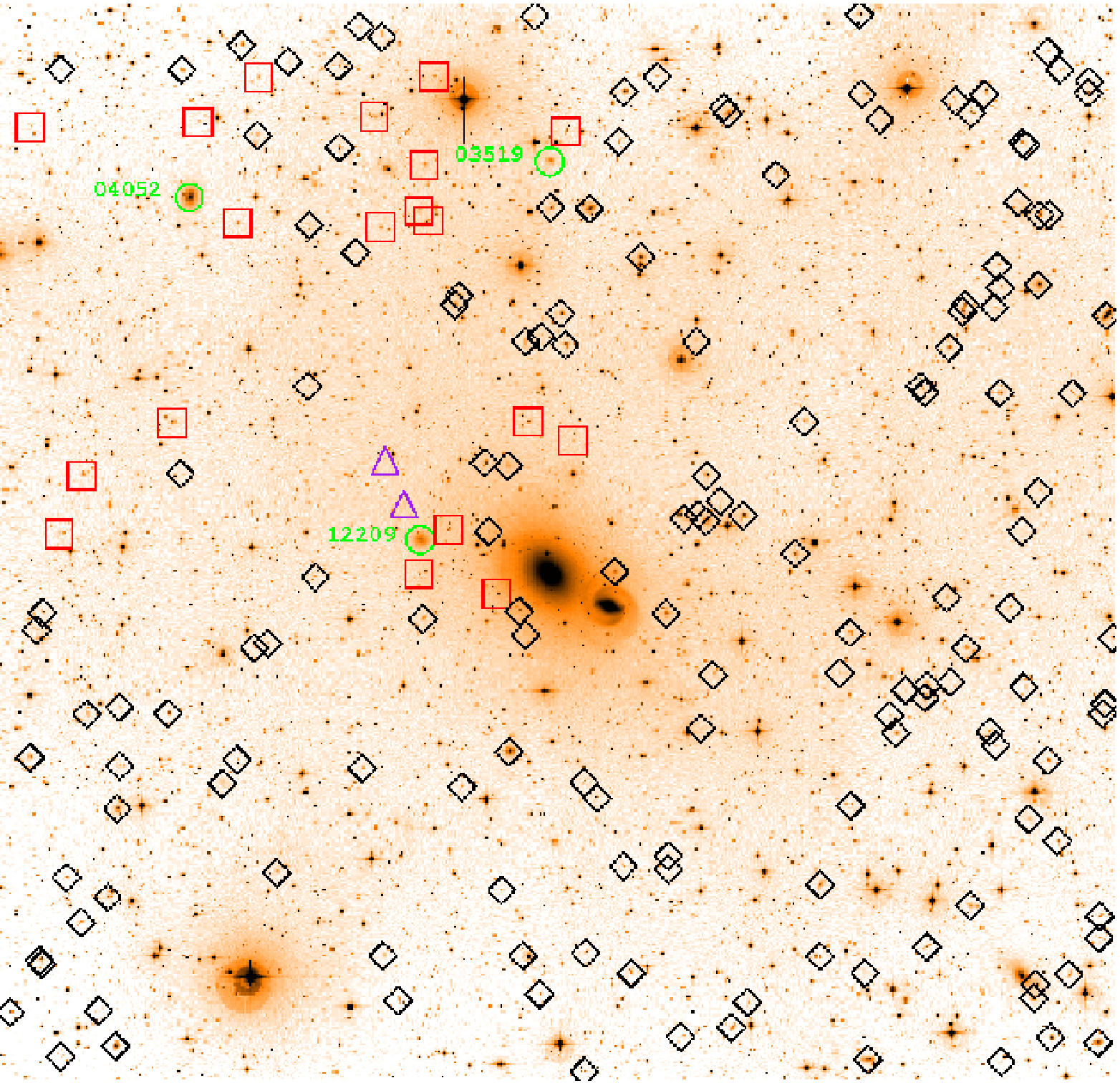}
\includegraphics[width=9.2truecm]{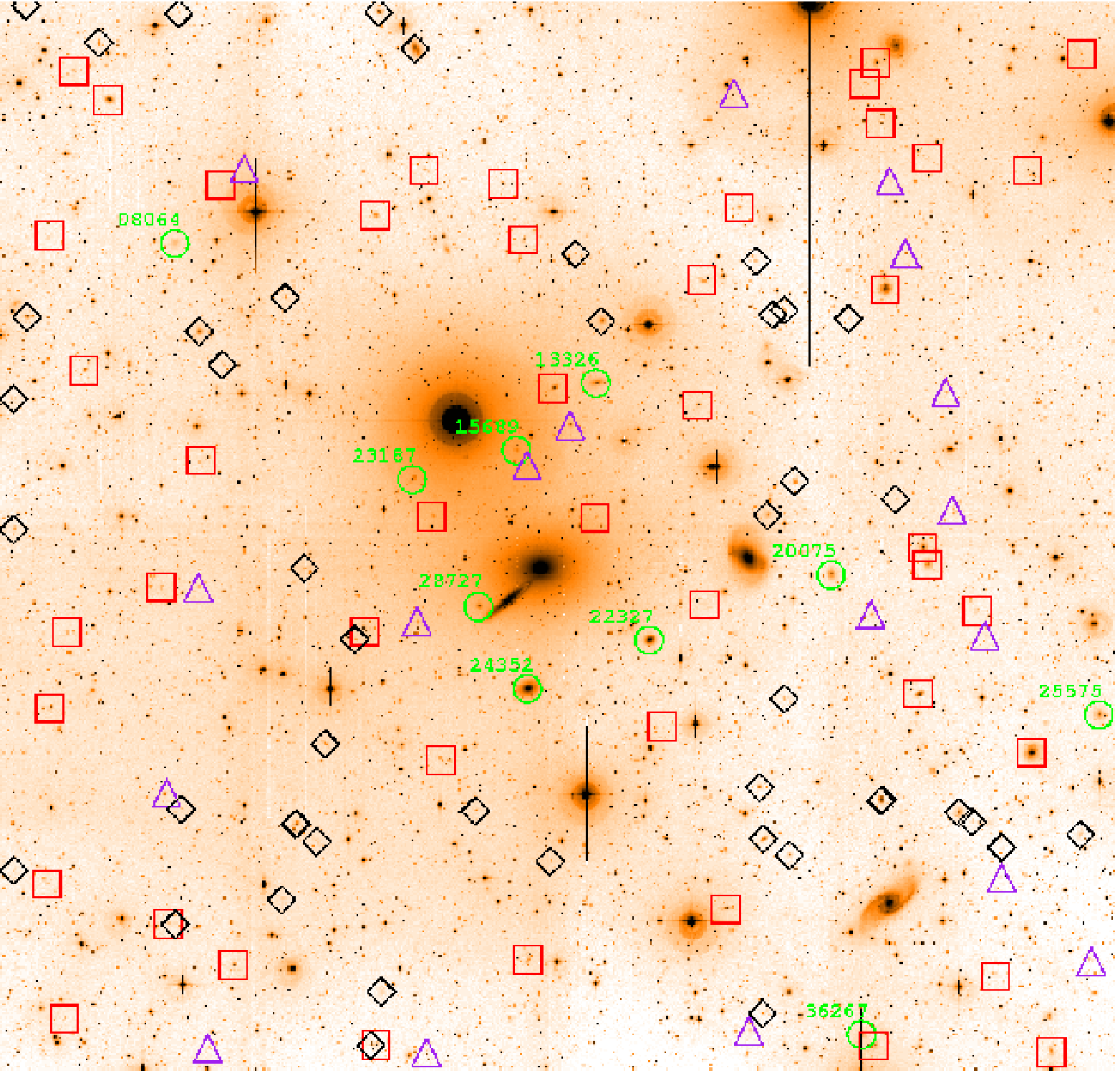}
\caption{Results of VIMOS observations superposed on the WFI images: RR~143 (top left), RR~210 (top right), RR~216 (bottom left), and RR~242 (bottom right). Each field is centred on the bright E galaxy of the pair. Due to an error in the service-mode observations, the field of RR~143 was not dithered and  thus shows the gaps between the single CCDs. Each WFI field was covered by 4 VIMOS pointings, one for each quadrant of the WFI field of view. However, not all four quadrants were observed for each group, due to incomplete service-mode observations. Marked with different symbols are: new group members (green circles), confirmed background galaxies (red squares), spectroscopically observed candidates too faint for redshift measurement (blue triangles), and spectroscopically non-observed objects (black diamonds). The newly identified group members are also labelled with their object ID (see Table~\ref{table3}). The different spectroscopic coverage of each field can be seen, since the not-covered quadrants only contain black diamonds.
\label{fig1}}
\end{figure*}

Each WFI field is covered by four VIMOS observing blocks, one VIMOS pointing for each quadrant of the WFI field of view. The observations of each single block were divided into two exposures.  Bias, flat-field, and standard-star calibration files were associated with each observing block as well as the helium-argon lamp spectrum for wavelength calibration. Observations were obtained in service mode to guarantee optimal observing conditions. Unfortunately, the complete four quadrants were obtained only for RR~143 and RR~242. Two quadrants were observed for RR~210 and only one for RR~216.  Table~\ref{table1} provides the observing log for the four E+S systems. Figure~\ref{fig1} shows the WFI fields with the results of our VIMOS observations.

The basic CCD reduction of each frame containing spectra as well as the wavelength calibration was completed using the ESO software environment {\tt
Esorex}.  The 2D spectrum of each slit was coadded to the corresponding one from the second exposure. The object(s) in each
slit was(were) then extracted into 1D object-spectra containing the total light of each target. Finally, each wavelength-calibrated spectrum was stored
as a single {\tt FITS}-file for further processing.

Redshifts were measured using the cross-correlation technique \citep[see e.g.,][]{Tonry79}. To provide a reliable estimate of radial velocities and
their uncertainties, 5 stellar template spectra were used.  The {\tt IRAF}\footnote{IRAF is distributed by the National Optical Astronomy
Observatories, which are operated by the Association of Universities for Research in Astronomy, Inc., under cooperative agreement with the National
Science Foundation.} package {\tt rvsao} provides the {\tt xcsao} task to measure radial velocities via cross-correlation.  During this interactive
cross-correlation procedure, the result of the cross-correlation was always directly inspected to avoid spurious results. In some cases, usually for emission-line dominated spectra, the lines could not be identified by {\tt xcsao} and were processed using the {\tt IRAF} task {\tt splot}. With this task a Gaussian fit to each spectral line can be performed. The adopted redshift is then the average of all fitted lines, and its error is given by the
standard deviation in the different redshift values.

\begin{table}
\begin{tiny}
\caption{Observation statistics of the candidate samples. }
\begin{tabular}{lcccc}
\hline\hline
   & RR~143 & RR~210 & RR~216 & RR~242\\
\hline
Candidate sample & 171 & 190 & 178 & 118 \\
Observed & 106 (62\%) & 55 (29\%)& 24 (13\%) & 73 (62\%)\\
Measured z total & 84 (79\%) & 52 (95\%) & 22 (92\%) & 55 (75\%) \\
Measured z {\tt xcsao} & 69 & 50 & 22 & 45 \\
Measured z {\tt splot} & 15 & 2 & -- & 10 \\
Members & 2 (2\%) & 2 (4\%) & 3 (14\%) & 10 (18\%) \\
Members corr. & 4.4 & 4.9 & 15 & 15.4 \\
\multicolumn{5}{l}{}\\
\multicolumn{4}{l}{\bf Total number of spectroscopically observed objects:} & \multicolumn{1}{l}{\bf 258} \\
\multicolumn{4}{l}{\bf Total number of measured redshifts:} & \multicolumn{1}{l}{\bf 213 (83\%)} \\
\hline
\multicolumn{5}{l}{}\\
\end{tabular}
\footnotetext{}{
Row 1: Number of galaxies in the candidate sample. \\
Row 2: Number of galaxies observed with VIMOS, in brackets the percentage of  spectroscopically observed galaxies out of the candidate sample.\\
Row 3: Number of galaxies for which a redshift could be measured, in brackets the percentage of galaxies with measured redshift out of all  spectroscopically observed galaxies.\\
Row 4: Number of redshifts obtained by cross-correlation with {\tt xcsao}. \\
Row 5: Number of redshifts obtained by fitting single lines with {\tt splot}. \\
Row 6: Number of newly discovered members, in brackets the percentage of newly found member galaxies out of all galaxies with measured redshifts.\\
Row 7: Completeness corrected number of members (see Section~\ref{results}).\\

}
\label{table2}
\end{tiny}
\end{table}

Table~\ref{table2} reports the statistics of the observational campaign. For the newly confirmed members of the E+S systems, we obtained 
accurate surface photometry from our WFI images.  The methods adopted are fully explained in Paper~III.

\section{Results}\label{results}

Redshift measurements allowed us to identify faint galaxies likely to be physically associated with each E+S pair system. However, the spectroscopic coverage of the fields around the four pairs is not uniform, due to incomplete service-mode observations. In spite of two approved observing programs, only $\sim 2/3$ of the total area (the 4 WFI fields) was observed with VIMOS. Additionally, the incompleteness differed between the 4 fields with RR~143 and RR~242 being covered completely, while for RR~210 and RR~216 only 50\% and 25\%, respectively, of their fields were covered.

\begin{figure*}
\begin{center}
\includegraphics[width=17cm]{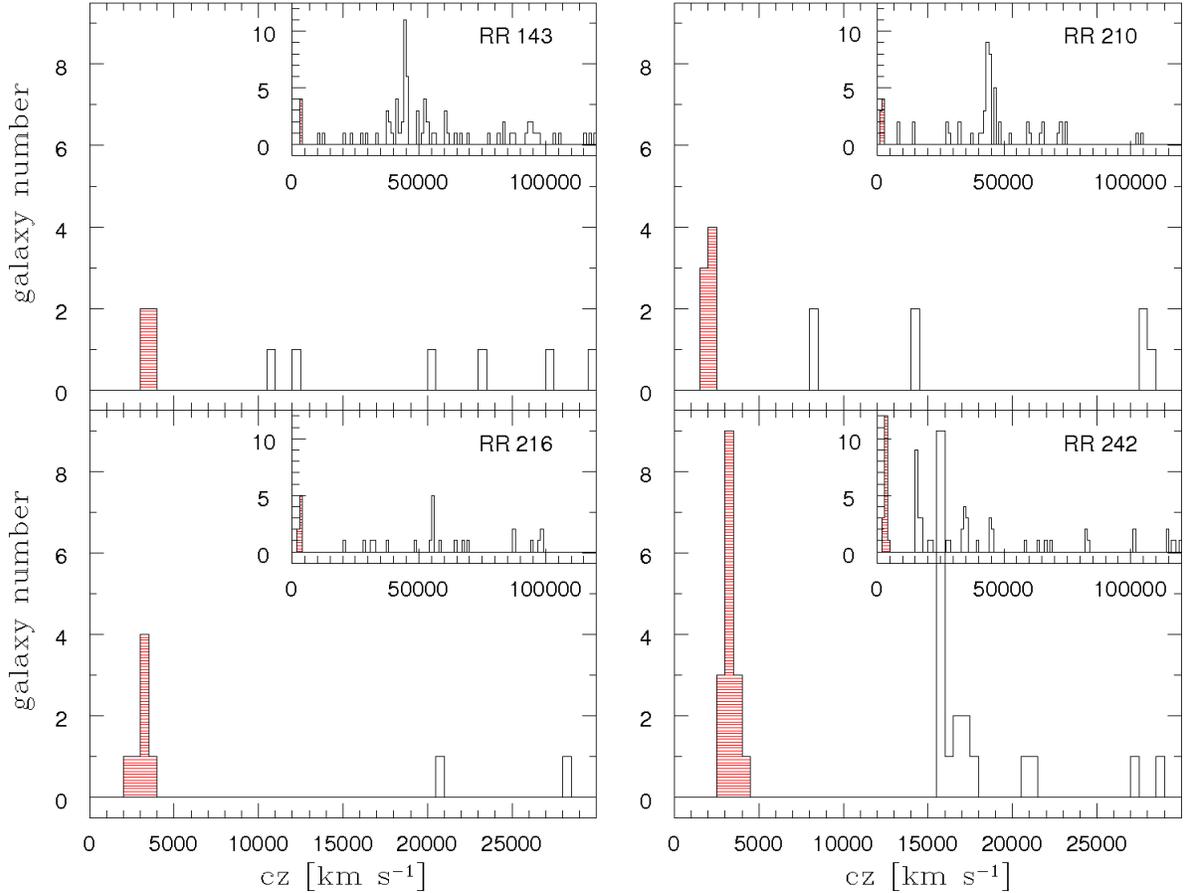}
\caption{Histogram of galaxy redshifts measured in the WFI fields. Large windows show the velocity range from 0 $-$ 30 000 km s$^{-1}$, i.e. up to
$\sim$ 10 times the group systemic velocity. The whole range of redshifts (up to $\sim$ 130 000 km s$^{-1}$) is shown in small windows in the upper right
corner of each panel. Galaxies considered to be group members (new and previously known, see text) are shaded in red. The bin size is 500 km s$^{-1}$.
\label{fig2}}
\end{center}
\end{figure*}

Figure~\ref{fig2} shows the distribution of measured radial velocities for each field. Each group panel shows the velocity range up to 30 000 km~s$^{-1}$ ($\sim$10 times the group velocity), whereas the full range of velocities found up to $\sim$ 130 000 km~s$^{-1}$ is plotted in the small window in the upper right corner of each group panel. Galaxies belonging to each group are shaded in red. To show the group structure in greater detail,we included the already known group members inside the WFI field of view in each histogram. RR~143 (top left panel) and RR~210 (top right panel) show prominent concentrations at $\sim$ 40 000 km~s$^{-1}$, while RR~242 (bottom right panel) shows a weaker concentration at $\sim$ 16 000 km~s$^{-1}$.
Galaxies in the field of RR~216 (bottom left panel) seem to show a less clustered distribution in redshift space. However, only one quadrant of the WFI field was observed with VIMOS in the latter case, resulting in a smaller number of galaxies with measured redshifts than in the other fields (see
Table~\ref{table2}).

\subsection{Identification of faint members and completeness correction}\label{completeness}

\begin{table*}
\centering
\begin{tiny}
\caption{Summary of the properties of the new members of the E+S systems.}
\label{table3}
\begin{tabular}{lccccccccccc}
\hline\hline
 ID & $\alpha$ & $\delta$ & v$_{rad}$ & R$^1$ & D$^2$ & m$_R$ & $(V-R)$ & n & $r_e$ & $\mu_0$ & $\mu_e$ \\
    & & & [km s$^{-1}$] & & [Mpc] & [mag] & [mag] & & [kpc] & \multicolumn{2}{c}{ [mag arcsec$^{-2}$]}\\
\hline
RR143\_09192 & 06 47 53.25 & -64 10 55.1 & $3296\pm46~$ & ~3.34 & 0.122 & 17.25 & 0.56 & 1.05 & 1.30 & 21.30 & 23.23 \\
RR143\_24246 & 06 47 07.47 & -64 29 15.8 & $3724\pm73~$ & ~3.87 & 0.269 & 16.95 & 0.34 & 1.57  &  1.40  &  19.73  &  22.80  \\
  & & & & & & & & & & & \\
RR210\_11372 & 12 07 43.28 & -29 43 10.9 & $1927\pm183$ & ~3.08 & 0.117 & 17.59 & 0.50 & 0.83 & 0.97 & 22.24 & 23.68 \\
RR210\_13493$^3$ & 12 06 50.26 & -29 36 23.5 & $2008\pm45~$ & ~9.08 & 0.080 & 14.26 & 0.52 & 1.98 & 1.30 & 18.03 & 21.97 \\
  & & & & & & & & & & & \\
RR216\_03519 & 12 25 17.80 & -39 33 45.6 & $3716\pm134$ & ~2.13 & 0.165 & 17.26 & 0.51 & 1.02 & 1.22 & 21.41 & 23.28 \\
RR216\_04052$^4$ & 12 26 11.30 & -39 34 47.1 & $3193\pm50~$ & ~9.63 & 0.209 & 14.57 & 0.60 & 1.07 & 2.52 & 20.18 & 22.15 \\
RR216\_12209 & 12 25 37.04 & -39 44 36.4 & $2697\pm62~$ & ~5.32 & 0.054 & 15.58 & 0.59 & 0.74 & 2.17 & 21.57 & 22.82 \\
  & & & & & & & & & & & \\
RR242\_08064 & 13 22 10.44 & -43 32 55.6 & $3441\pm60~$ & ~2.58 & 0.202 & 17.85 & 0.64 & 1.21 & 1.32 & 21.88 & 24.15 \\
RR242\_13326 & 13 21 03.88 & -43 36 58.1 & $3172\pm45~$ & ~3.42 & 0.080 & 16.15 & 0.60 & 1.42 & 2.11 & 20.15 & 22.87 \\
RR242\_15689 & 13 21 16.54 & -43 38 53.2 & $3731\pm81~$ & ~7.20 & 0.050 & 17.73 & 0.74 & 1.38 & 0.73 & 20.47 & 23.11 \\
RR242\_20075 & 13 20 26.71 & -43 42 27.5 & $3045\pm52~$ & ~6.63 & 0.120 & 16.33 & 0.55 & 1.51& 1.38 & 20.02 & 22.95 \\
RR242\_22327$^5$ & 13 20 55.54 & -43 44 20.2 & $3237\pm70~$ & ~9.63 & 0.054 & 15.06 & 0.69 & 1.34 & 0.98 & 18.38 & 20.94 \\
RR242\_23187 & 13 21 33.01 & -43 39 43.0 & $3400\pm88~$ & ~0.00 & 0.058 & 17.37 & 0.44 & 0.75 & 0.95 & 20.54 & 21.82 \\
RR242\_24352$^6$ & 13 21 14.82 & -43 45 43.2 & $2697\pm45~$ & 10.05 & 0.050 & 13.71 & 0.62 & 2.35 & 1.68 & 16.00 & 20.76 \\
RR242\_25575 & 13 19 44.25 & -43 46 28.0 & $2655\pm84~$ & ~4.26 & 0.239 & 16.42 & 0.60 & 2.24 & 1.73 & 19.02 & 23.53 \\
RR242\_28727 & 13 21 22.55 & -43 43 21.6 & $3017\pm87~$ & ~6.80 & 0.030 & 17.05 & 0.64 & 1.16 & 0.69 & 20.08 & 22.25 \\
RR242\_36267 & 13 20 21.80 & -43 55 38.4 & $2997\pm40~$ & ~3.19 & 0.245 & 18.67 & 0.50 & 0.89 & 1.07 & 22.32 & 23.90 \\
\hline
\multicolumn{12}{l}{}\\
\multicolumn{12}{l}{$^1$ confidence parameter R from the cross-correlation procedure. If R = 0 then the lines were measured by hand with {\tt splot}.} \\
\multicolumn{12}{l}{$^2$ projected distance from the E member of the pair.} \\
\multicolumn{12}{l}{$^3$ RR210\_13493 = 2MASX~J12065029-2936236.} \\
\multicolumn{12}{l}{$^4$ RR216\_04052 = 2MASX~J12261133-3934474.} \\
\multicolumn{12}{l}{$^5$ RR242\_22327 = ESO~270-~G~001.} \\
\multicolumn{12}{l}{$^6$ RR242\_24352 = ESO~270-~G~003.} \\
\end{tabular}
\end{tiny} 
\end{table*}


Different membership criteria are considered in the literature. A galaxy is often considered to be a member of a structure if the velocity difference
between the galaxy and the structure's systemic velocity is lower than a certain value. For instance, \citet{Karachentsev90} and \citet{Hickson92} adopted $|(v_{galaxy} - v_{group})| \leq$ 1000 km~s$^{-1}$, while \citet{Ramella94} used $\leq$ 1500 km~s$^{-1}$. 
Group membership may also be defined in terms of the group velocity dispersion, $\sigma_{group}$, leading to a selection that is more capable of being adapted to the true group's properties. In this case, a limit of $|(v_{galaxy} - v_{group})| \leq 3\sigma_{group}$ has been used, reflecting the approximate dynamical boundaries of the group \citep[see e.g.,][]{M00,Cellone05,Firth06}.
Different membership criteria applied to our sample yield the same result: there are no galaxies close to the velocity limits set by the above criteria (the group velocity dispersions used in the flexible group membership criterion are listed in Table~\ref{table5}). Line 6 of Table~\ref{table2} gives the number of newly identified members for each of our groups. Only a small fraction of the candidates turned out to be new members of our four E+S systems. However, the E+S systems are clearly defined structures in redshift space suggesting that they are real, albeit sparse, physical associations. Coordinates, redshifts with estimated uncertainty, projected distance from the bright E, total R-band magnitude, and $(V-R)$ colour of the new members are presented in Table~\ref{table3}. The structural parameters (Sersic index $n$, effective radius $r_e$, and central and effective surface brightness $\mu_0$ and $\mu_e$) given in the rightmost 4 columns of Table~\ref{table3} are obtained from one-component Sersic-model fits completed with GALFIT \citep[][see Paper~III]{Peng02}.

\begin{figure}
\includegraphics[width=9cm]{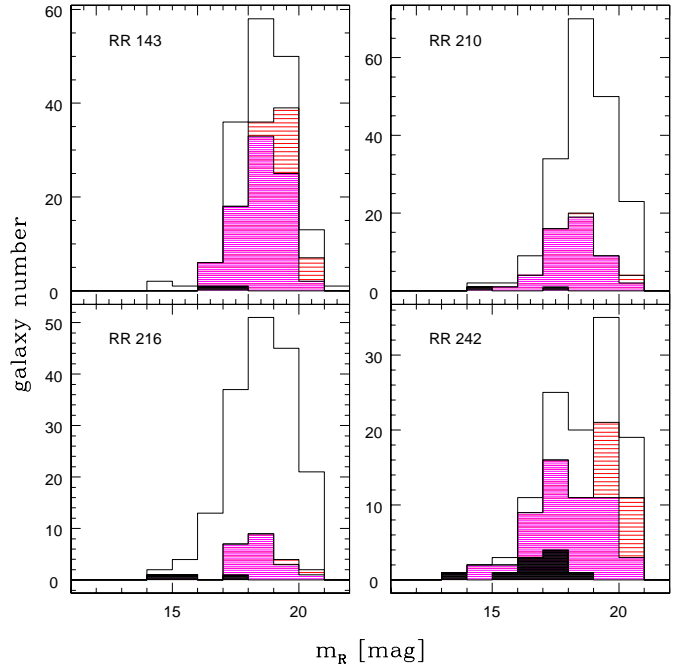}
\caption{Distribution of R-band magnitudes in all 4 group samples: RR~143 (top left), RR~210 (top right), RR216 (bottom left) and RR~242 (bottom right). The different histograms show the full candidate sample (unshaded),  spectroscopically observed galaxies (red, mildly shaded), galaxies with measured redshift (magenta, shaded), and new member galaxies (black, heavily-shaded). They correspond to the samples given in lines 1, 2, 5, and 6 in Table~\ref{table2}. The bin size is 1 magnitude.
\label{fig3}}
\end{figure}

Figure~\ref{fig3} shows the distribution of R-band magnitudes for the candidate list (Paper~III), for all objects observed spectroscopically (red), all galaxies with measured redshifts (magenta), and for those adopted as group members (black). Both RR~143 and RR~242, with complete VIMOS pointings, show a reasonable degree of completeness. In contrast, for both RR~210 (2 pointings) and especially RR~216 (1 pointing) the number of candidates without spectroscopy is very high (see also Table~\ref{table2}). In addition to the missing pointings in RR~210 and RR~216, with 50\% and 25\% coverage, respectively, we must also consider two additional sources of incompleteness affecting the number of spectroscopically observed candidates. The major source of incompleteness is caused by instrumental constraints: (a) the gaps between the 4 CCDs in the VIMOS field of view reduce the analysed area by about 22\%, and (b) the HR\_blue grism allows only one slit to be placed along the dispersion direction, i.e., galaxies with similar declinations (closer than the slit length) cannot be observed with a single pointing. The second type of incompleteness depends on the source magnitude, i.e., redshifts of fainter objects become increasingly difficult to measure with the adopted exposure times. The magnitude-dependent incompleteness starts at R $\sim$ 18.5~mag, while at brighter magnitudes the incompleteness is determined by the instrumental constraints. Therefore, a magnitude-limited completeness correction was adopted.

\begin{figure}
\includegraphics[width=9cm]{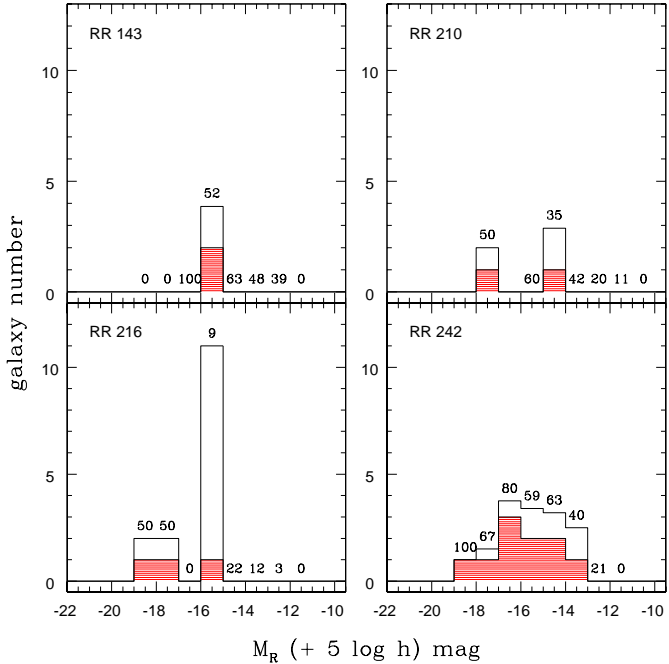}
\caption{Completeness correction of the number of member galaxies in each bin of absolute magnitude $M_R$. The bin size is 1 magnitude. The true
number of identified members per magnitude bin is shaded in red, whereas the number of expected members is plotted in white. The numbers over each bin
represent the percentage of candidates that have a measured $z$ in each magnitude bin, i.e., the spectroscopic completeness in that bin. If there is no number above a bin, then no candidates are located in this bin.
\label{fig4}}
\end{figure}

Our photometric selection criteria (see Section~\ref{obsandred} ) may also bias the number of faint member galaxies. When selecting the candidate objects, our first goal was to complete reliably the surface photometry to check if their structural properties are consistent with them being faint galaxies associated with their respective pair. To obtain a reliable estimate of the surface photometric parameters, the galaxy size (or the effective radius) should exceed the size of the seeing disk, which has a $FWHM < 1^{\prime\prime}$ in all images (see Table~4 in Paper~III). The detection size $a$ given by {\tt SExtractor} is not directly related to the galaxy's effective radius or the FWHM, but it gives the semi-major axis length of the detection ellipse, which is most likely larger than the galaxy's effective radius (which contains only half of the light). To ensure that we selected galaxies with effective radii larger than the area affected by the seeing, we chose a generous detection size limit of $1.5^{\prime\prime}$, which corresponds to a physical size of $\sim$ 400 pc at the distance of RR~242. This limit is also reasonable in a physical sense, since it is smaller than the smallest Local Group dwarf galaxies found by \citet{Mat98}, and reaches the domain of dwarf spheroidal galaxies (dSph), which are typically smaller than 500 pc.
Since small galaxies tend to be faint, the size limit leads to incompleteness of the candidate sample at faint magnitudes. Plotting detected size against magnitude, we found that this incompleteness starts at $m_R \sim 18$. This is also visible in Fig.~\ref{fig3}, where the magnitude histogram of the candidate sample bends over at around this magnitude. This incompleteness, however, does not affect galaxies at brighter magnitudes, since we did not exclude any objects brighter than $m_R \sim 18$ via this size cut-off. This apparent magnitude corresponds to an absolute magnitude of $M_R \sim -15 +5 \log h_{100}$, which is the limit to which we compare our results with the literature in the discussion (also because the spectroscopic completeness is higher than 50\% for brighter magnitudes, see below). Therefore, we can conclude that our size cut-off does not affect the results we present here.\\

\begin{figure}
\includegraphics[width=9cm]{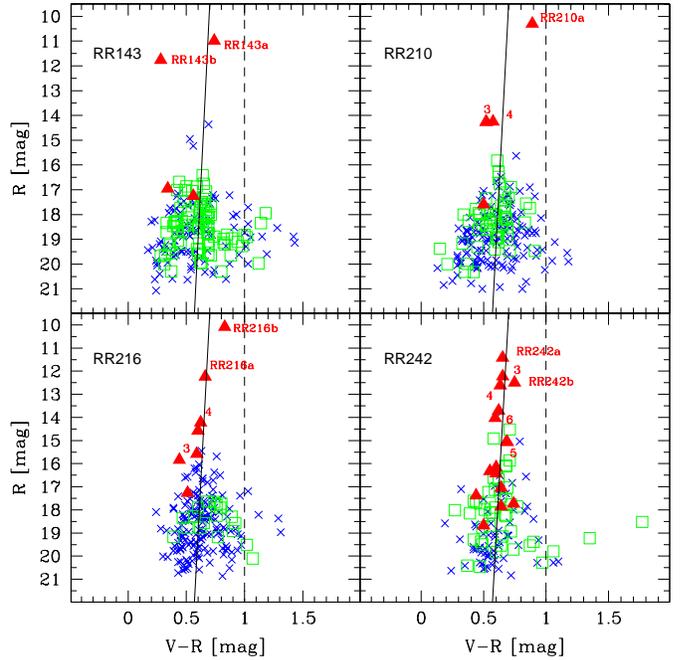}
\caption{Colour-magnitude relation for the candidate samples in the fields of the 4 E+S systems. Members are plotted as red triangles, previously known group members are additionally labelled with their ID (see Fig.~6-9 in Paper~III). Spectroscopically observed candidates that were found to be in the background are plotted as green squares, while candidates without measured $z$ are plotted as blue crosses. The solid line is a fit to the red-sequence of the Virgo-Cluster \citep{Vis77}, whereas the dashed vertical line represents the colour cut-off applied to the candidate sample.
\label{fig5}}
\end{figure}

We adopt a simple completeness-correction criterion based on the assumption that the same fraction of members is present in the sample of  spectroscopically observed and not observed objects.  The fraction of confirmed members in the sample with measured redshifts is computed for each magnitude bin and multiplied by the number of galaxies in the candidate sample found in the respective bin. This yields the total completeness-corrected number of members in each magnitude bin. The sum over all magnitude bins then gives the corrected number of group members given in line 7 of Table~\ref{table2}. 
Figure~\ref{fig4} shows the completeness-corrected number of members in each magnitude bin (white histogram) with the true number of confirmed members indicated in red. The number printed above each bin is the percentage of the spectroscopic completeness in the respective bin. So ``100'' means that the completeness is 100\% and all candidates in this bin have a measured redshift. If there is no number above a bin then this bin is empty, i.e., there are no candidates within this magnitude range.
\begin{table*}
\begin{tiny}
\begin{center}
\caption{Results of the Bulge-disk decomposition.}
\label{table4}
\begin{tabular}{lcccccccccc}
\hline\hline
Object & Morph. type & \multicolumn{2}{c}{de Vaucouleurs bulge} & \multicolumn{3}{c}{Sersic bulge} & B/T$_{deV}$ & B/T$_{S}$ & $\Delta$ B/T & $\Delta \chi ^2_\nu$ \\
     & type & bulge & disk & bulge & bar & disk & & & & \\
     & & $m_{R}$ & $m_{R}$ & $m_{R}$ & $m_{R}$ & $m_{R}$ & & & & \\
\hline
RR143\_09192 & SB0 & 19.10 & 17.33 & 17.31 & 20.70 & 18.55 & 0.16 & 0.77 & -0.6 & 30.6\\
RR143\_24246 & S0 &18.36 & 17.37 & 18.47 & ... & 17.51 & 0.29 & 0.29 & -0.01 & 0.9\\
 & & & & & & & & & \\
RR210\_11372 & dE & ... & 17.58 & ... & ... & 17.58 & 0.00 & 0.00 & 0 & 0.36\\
RR210\_13493 & SB0 & 15.11 & 14.87 & 16.18 & 17.30 & 14.53 & 0.44 & 0.23 & 0.22 & 10.6\\
 & & & & & & & & & \\
RR216\_03519 & dE & 21.95 & 17.29 & 20.69 & ... & 17.32 & 0.01 & 0.04 & -0.03 & 0.1\\
RR216\_04052 & SB0 & 17.95 & 14.67 & 17.39 & ... & 14.70 & 0.05 & 0.08 & -0.03 & 1.5\\
RR216\_12209 & Sc & 22.07 & 15.63 & 21.79 & ... & 15.63 & 0.00 & 0.00 & 0 & 0.1\\
 & & & & & & & & & \\
RR242\_08064 & dE & 19.25 & 18.18 & 19.28 & ... & 18.30 & 0.27 & 0.29 & -0.02 & 0.3\\
RR242\_13326 & S0 & 17.17 & 16.54 & 18.64 & ... & 16.29 & 0.36 & 0.10 & 0.26 & 0.1\\
RR242\_15689 & dE & 19.52 & 18.10 & 20.93 & ... & 17.95 & 0.21 & 0.06 & 0.15 & 0.3\\
RR242\_20075 & dE & 17.39 & 16.79 & 18.74 & ... & 16.53 & 0.37 & 0.12 & 0.25 & 0.8 \\
RR242\_23187 & S & ... & ... & 17.63 & ... & 17.96 & ... & 0.57 & ... & ... \\
RR242\_22327 & SB0 & 16.72 & 15.51 & 17.58 & 16.56 & 15.79 & 0.25 & 0.41 & -0.16 & 2.2\\
RR242\_24352 & SB0 & 14.59 & 14.52 & 15.31 & 17.36 & 14.17 & 0.49 & 0.29 & 0.2 & 21.4\\
RR242\_25575 & SB? & 16.27 & 17.70 & 17.62 & ... & 16.77 & 0.79 & 0.31 & 0.47 & 0.8\\
RR242\_28727 & dE & 18.25 & 17.38 & 17.95 & ... & 17.40 & 0.31 & 0.38 & -0.07 & -82.9\\
RR242\_36267 & dE & 21.31 & 17.92 & 23.06 & ... & 17.89 & 0.04 & 0.01 & 0.03 & 22.5 \\
\hline
\multicolumn{11}{l}{}\\
\multicolumn{11}{l}{Notes: Bulge and disk magnitudes and bulge-to-total light (B/T) ratios of the fit with de Vaucouleurs bulges (columns 3-4)} \\
\multicolumn{11}{l}{ and Sersic bulges (columns 5-7) respectively. Columns 8-10 give the Bulge-to-Total light ratios of the two models respectively}\\
\multicolumn{11}{l}{and their difference. The last column gives the difference between the $\chi ^2_\nu$ of the de Vaucouleurs and the Sersic-model fit.}
\end{tabular}
\end{center}
\end{tiny}
\end{table*}

Poor statistics in the RR~216 field makes the application of the above criterion very uncertain. We note, e.g., the high number of expected members in the $-15 > M_R > -16$ magnitude bin: the only candidate with measured redshift in this bin was confirmed as a member. Following the above criterion, all galaxies in the candidate sample at this magnitude were assumed to be members. However, comparing the number of confirmed members with the number of galaxies with measured redshifts (Table~\ref{table2}) also suggests a higher number of member galaxies for RR~216, approaching a number similar to that of the X-ray bright RR~242.\\

Incompleteness effects are certainly an issue in our sample, although they are not expected to play a major role for absolute magnitudes as faint as $M_R \sim -17 +5 \log h_{100}$. Due to instrumental constraints, our spectroscopy missed 3 ``bright" candidates in RR~143, visible as the 2 first bins labelled with ``0'' in Fig.~\ref{fig4}. In all the other groups, the candidates brighter than $M_R = -17 +5 \log h_{100}$ without measured redshift are accounted for in the completeness correction: 1 object in RR~210 (bin labelled with ``50", 1 object added by correction), 2 objects in RR~216 (2 bins labelled with ``50", 2 objects added by correction) and 1 object in RR~242 (bin labelled with ``67", 1 object added by correction). Figure~\ref{fig4} also shows that, apart from RR~216 (where only one quarter of the field was covered), the spectroscopic completeness is above 50\% in all magnitude bins down to $M_R = -15 +5 \log h_{100}$. Any information fainter than this magnitude is not used in the comparison of our results with the literature and does not affect our conclusions.

\subsection{The photometric and structural properties of faint members}

In general, we find that the confirmed companions tend to be of intermediate luminosity, which is a domain populated by faint S0, spirals, and dwarf elliptical galaxies.

Figure~\ref{fig5} shows the colour-magnitude relation of each group. Confirmed member galaxies are indicated by red triangles. The solid line represents a fit to the red sequence of the Virgo Cluster \citep{Vis77} shifted to the pairs' distance, while the dashed line shows the colour cut-off applied to the candidate sample.
Galaxies follow the red sequence for early-type galaxies even at faint magnitudes. The new members have very uniform colours with no blue star-forming dwarfs found in our sample. This might be partially caused by the selection criteria used to construct the candidate sample, especially by the size cut-off. As discussed in section~\ref{completeness}, this size cut-off leads to the loss of galaxies below $m_R \sim 18$ (corresponding to $M_R \sim -15 +5 \log h_{100}$ at the farthest pair's distance). However, also above that magnitude, where the candidate samples are complete (photometrically), they do not contain blue galaxies. Blue galaxies fainter than $m_R \sim 18$, are not abundant in our sample, but those observed spectroscopically were found to be in the background.

\begin{figure*}
\resizebox{18cm}{!}{
\includegraphics[]{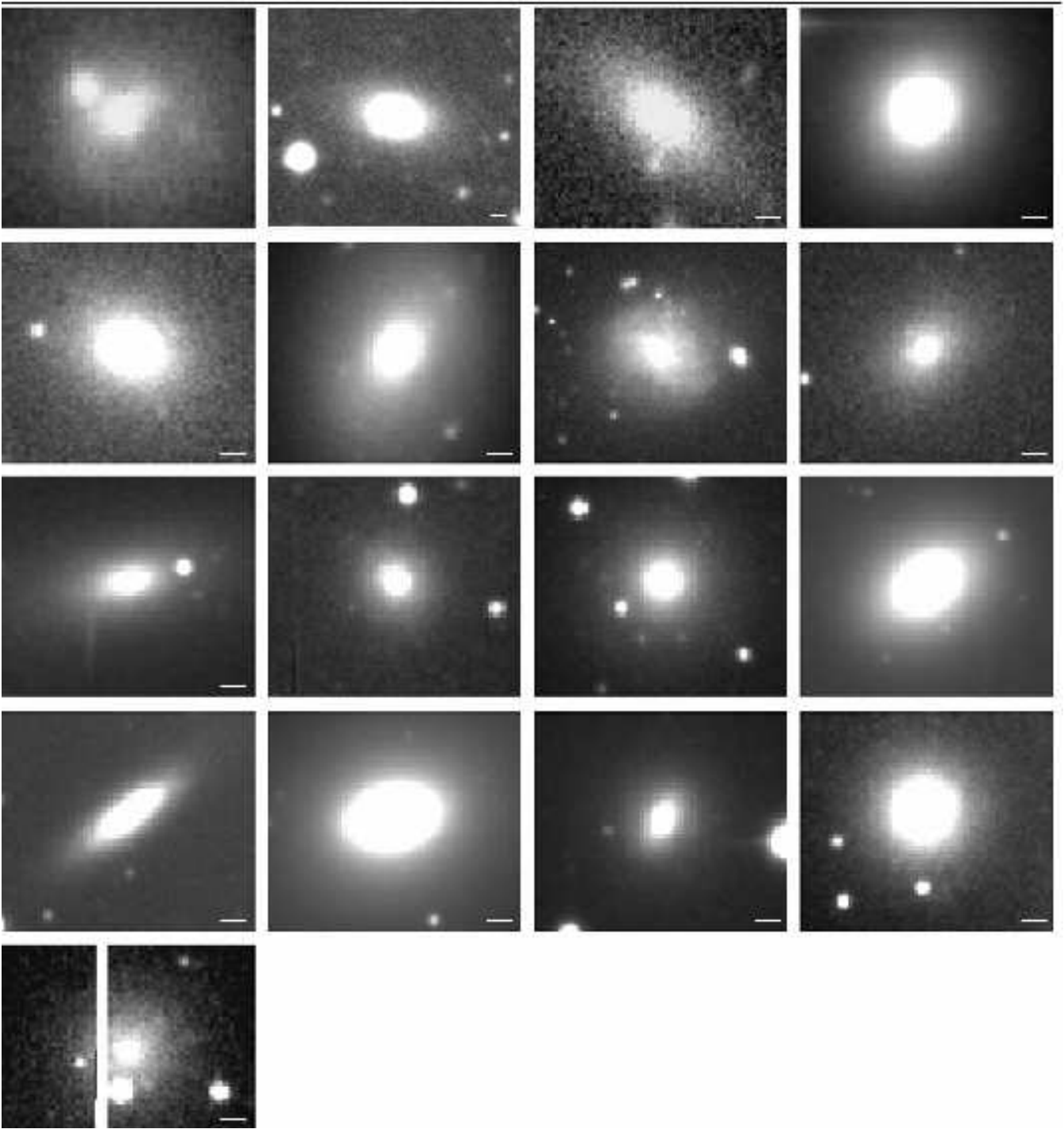}
}
\caption{R-band thumbnails of member galaxies of all 4 groups. From left to right: 1$^{st}$ row from the top: RR143\_09192, RR143\_24246, RR210\_11372, and RR210\_13493; 2$^{nd}$ row: RR216\_03519, RR216\_04052, RR216\_12209, and RR242\_08064; 3$^{rd}$ row: RR242\_13326, RR242\_15689, RR242\_20075, and RR242\_22327; 4$^{th}$ row: RR242\_23187, RR242\_24352, RR242\_25575, and RR242\_28727; last row: RR242\_36267. The scale bar has a length of 2 arcseconds in each image.
\label{fig6}}
\end{figure*}

In the following, we consider only galaxies identified as group members according to the redshift measurements. Figure~\ref{fig6} shows R-band images of new confirmed members. Galaxy morphologies were investigated with {\tt ELLIPSE} and {\tt GALFIT} (see Paper~III for a full explanation).  Results are compared with the galaxy morphologies of the \citet{ZM00} X-ray detected groups discussed in \citet{Tran01}.

\begin{figure*}
\resizebox{18cm}{!}{
\includegraphics[]{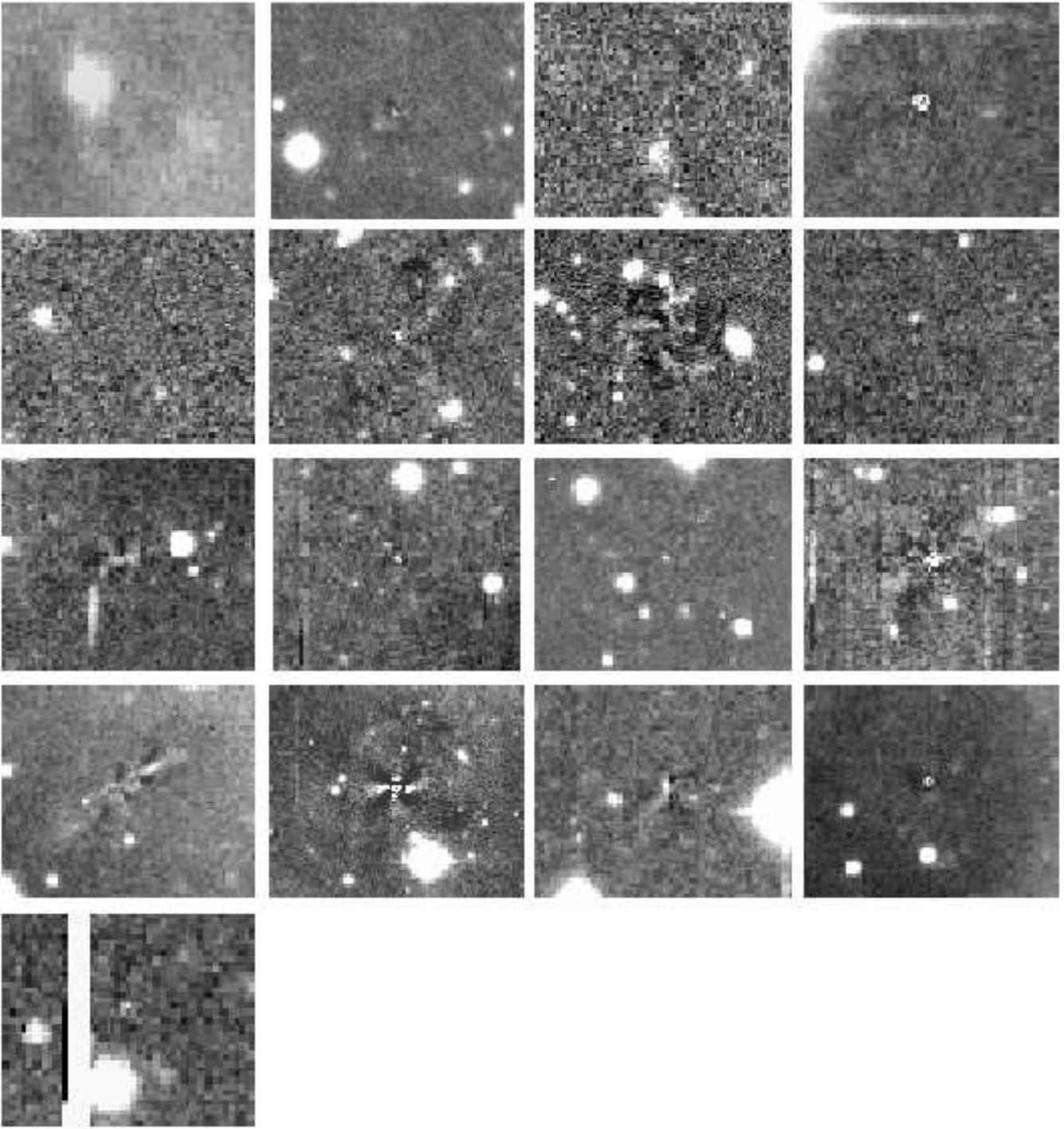}
}
\caption{Residual images after subtraction of a model constructed from the isophotal fit with {\tt ELLIPSE}. Objects are shown in the same order and 
with the same scale as in Fig.~\ref{fig6}. 
\label{fig7}}
\end{figure*}

\begin{figure*}
\begin{center}
\includegraphics[width=16cm]{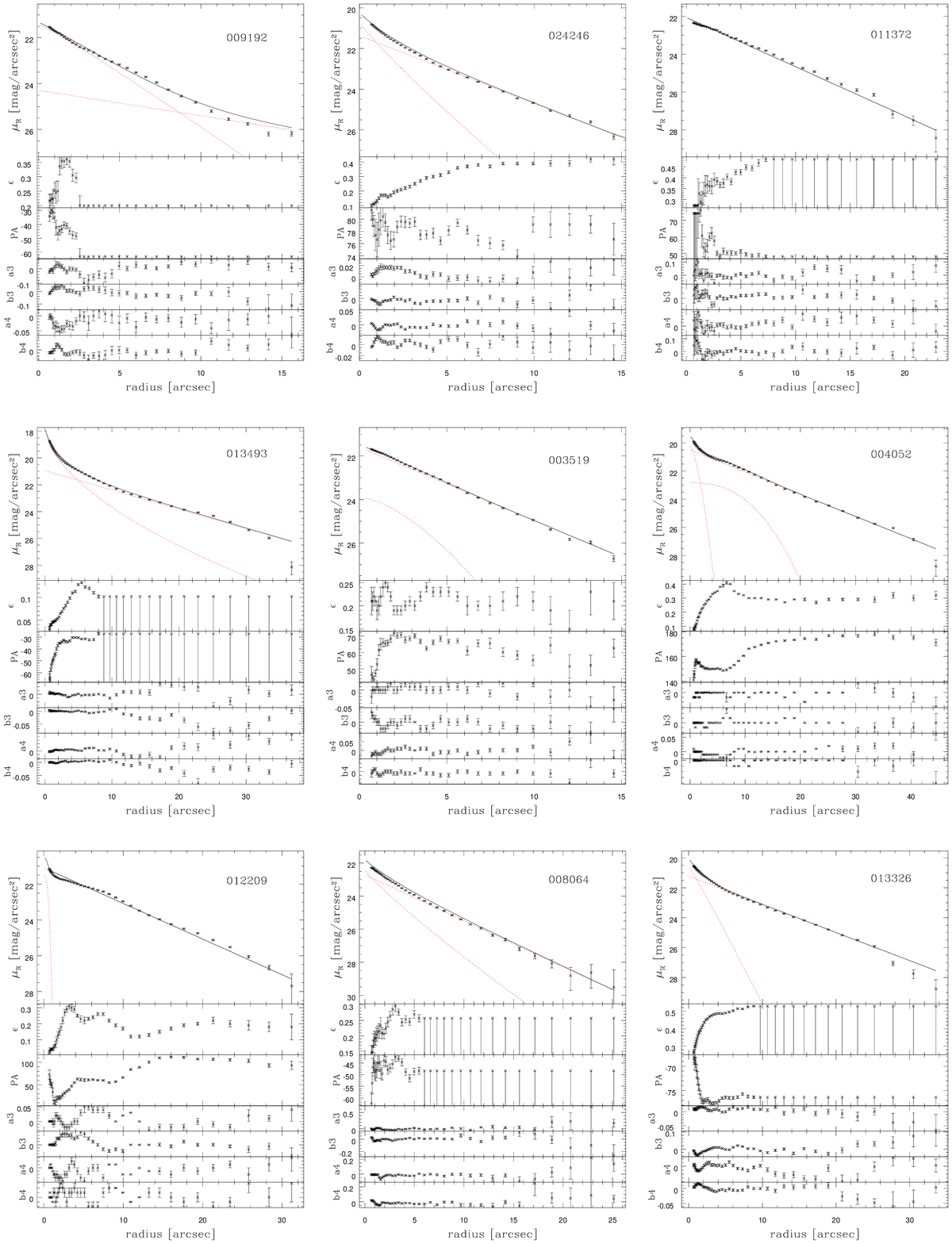}
\vspace{-0.5cm}
\caption{Surface photometry. The radial profiles of surface brightness, ellipticity, position angle, and higher order coefficients of the Fourier expansion $a_3$, $b_3$, $a_4$, $b_4$ of faint member galaxies in the R-band obtained by {\tt ELLIPSE}. Each successive isophote is plotted as a small cross, the error bars are the errors given by {\tt ELLIPSE}. The surface-brightness panel (top panel in each plot) also shows the profile of the 2-component model obtained from {\tt GALFIT}. The bulge and disk components are plotted as red dotted lines. If a third component (bar) improved the fit, it is also plotted. The total 2 (or 3) component galaxy model is then plotted as solid black line. Objects shown are RR143\_09192, RR143\_24246, RR210\_11372, RR210\_13493, RR216\_03519, RR216\_04052, RR216\_12209, RR242\_08064, and RR242\_13326.
\label{fig8}}
\end{center}
\end{figure*}

\begin{figure*}
\begin{center}
\includegraphics[width=16cm]{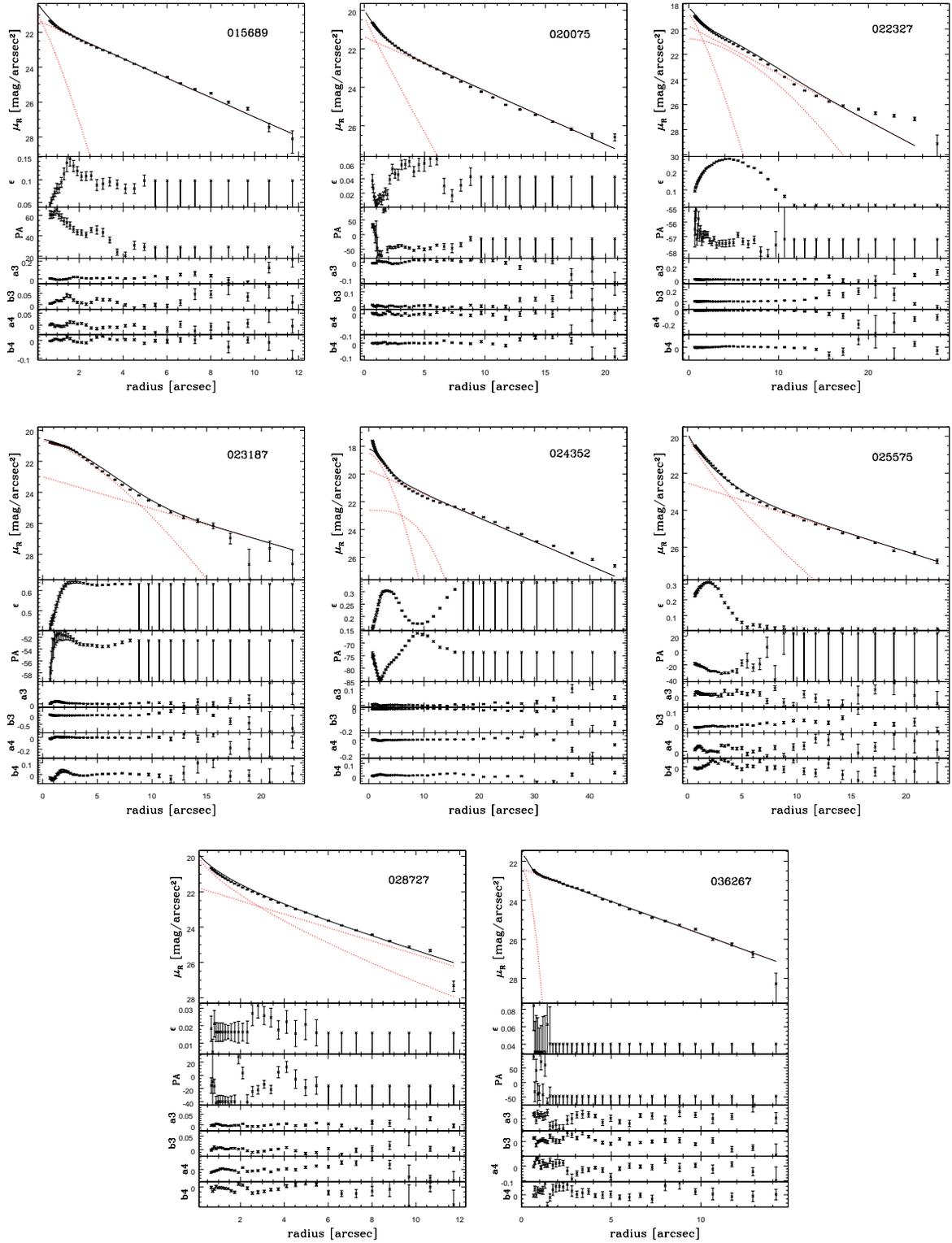}

\caption{Surface photometry as in Fig.~\ref{fig8}. Objects shown are RR242\_15689, RR242\_20075, RR242\_22327, RR242\_23187, RR242\_24352,
RR242\_25575, RR242\_28727, and RR242\_36267. 
\label{fig9}}
\end{center}
\end{figure*}

\begin{figure}
\includegraphics[width=9cm]{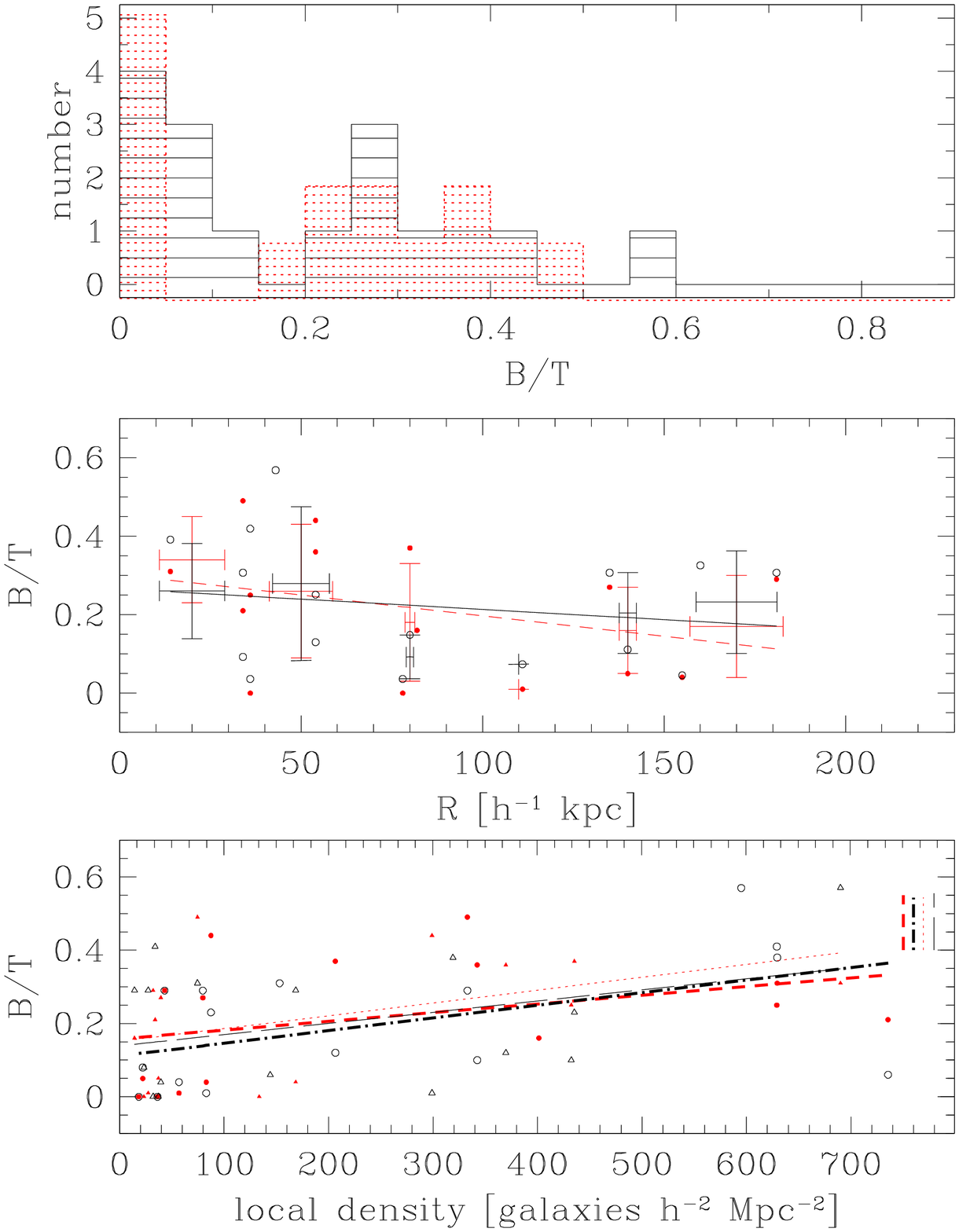}
\caption{Bulge-to-total light ratios (B/T) of member galaxies of the 4 groups. Black and open symbols indicate B/T obtained from fitting a Sersic bulge, red and solid symbols indicate the de Vaucouleurs bulge. {\it Top:} histogram of B/T ratios for the fit with Sersic bulges (solid black line) and de Vaucouleurs bulges (red dotted line). {\it Centre:} B/T versus radius from the pair elliptical. Lines are least squares fits to the data (black solid -- Sersic: red dashed -- de Vaucouleurs). {\it Bottom:} B/T versus local projected density. Values are computed with the closest 5 galaxies ($d_5$, circles) and the closest 3 galaxies ($d_3$, triangles). Lines are least squares fits to $d_5$ (dashed-dotted -- Sersic; red dashed -- de Vaucouleurs) and $d_3$ (long-dashed -- Sersic; red short-dashed -- de Vaucouleurs). The vertical lines in the upper right corner represent the rms scatter of each fit.
\label{fig10}}
\end{figure}

In Fig. 7, we display residual images after subtraction of a galaxy model constructed from the isophotal fit  with {\tt ELLIPSE}. Different galaxy substructures are clearly visible in the residual images including: asymmetries (RR143\_09192, RR242\_23187, RR242\_25575), bars (RR143\_09192, RR216\_04052,
RR242\_22327), filaments (RR242\_13326), and shells (RR242\_24352). The system of shells in RR242\_24352 extends to a radius of $\sim$ 6 $h_{100}^{-1}$ kpc and is not aligned ($\Delta$PA$\sim$30$^\circ$) with the semi-major axis of the galaxy. Different formation scenarios for stellar shells have been proposed including weak interactions, accretion of companions and major/minor mergers \citep[see e.g.,][]{Dupraz86,Hernquist87a,Hernquist87b}. In any case, they are considered clear evidence of environmental influence on galaxy evolution, which is apparently found even in rather sparse groups such as those in our sample.

Radial profiles of surface brightness ($\mu$), ellipticity ($\epsilon$), position angle (PA), and higher order coefficients of the Fourier expansion $a_3$, $b_3$, $a_4$, and $b_4$ are shown for each galaxy in Figs.~\ref{fig8} and \ref{fig9}. Isophotal profiles and the {\tt GALFIT} two-component models (see next paragraph) are shown. Dotted, red lines are Sersic (bulge) and exponential (disk) models, while solid lines represent the resultant total galaxy model. The figures show that the two-component models yield good fits to the real profiles in most cases. There are some residuals for RR242\_22327 and RR242\_24352, which exhibit very complex structures and are not well represented even by 3 components. The deviations from the model in the outskirts of RR242\_22327 are caused by an underestimated ellipticity of the outer isophotes. The ellipticity had to be fixed because of the high percentage of masked area due to some brighter foreground stars in the field. This was also the case for RR143\_09192: a foreground star close to the galaxy centre led to the loss of a considerable part of the galaxy image. This galaxy is not included in the B/T analysis in Fig.~\ref{fig10}.

The morphological classification listed in Table~\ref{table4} was completed using the surface photometric profiles in Figs.~\ref{fig8} and \ref{fig9}. The presence of bulge and disk components is clearly evident as a bend in the surface-brightness profile at the transition between the two components, due to the different shapes of the radial surface-brightness profiles of spheroids and disks. The surface brightness $\mu$ declines with radius as $\mu \sim r^{1/n}$. Disks (and dwarf ellipticals) have an exponential profile with $n \sim 1$, whereas spheroids are characterised by a higher value of $n$. Additionally, a disk is characterised by a constant $\epsilon$ and PA, whereas along the bulge, both $\epsilon$ and PA can vary. In this way, ellipticals (being pure bulges) and lenticular galaxies (having a bulge and a disk) can be distinguished easily.
Bars can also be identified in the surface photometric profiles, showing a constant surface brightness and PA in combination with a high $\epsilon$.
We find a number of bars in our faint galaxy sample. In 5/17 objects a bar is clearly distinguishable in the radial profiles. A weak bar is also suspected in (RR242\_25575). Bars can also be an indication of on-going interaction as shown by \citet{No87}.

\begin{figure}
\includegraphics[width=9cm]{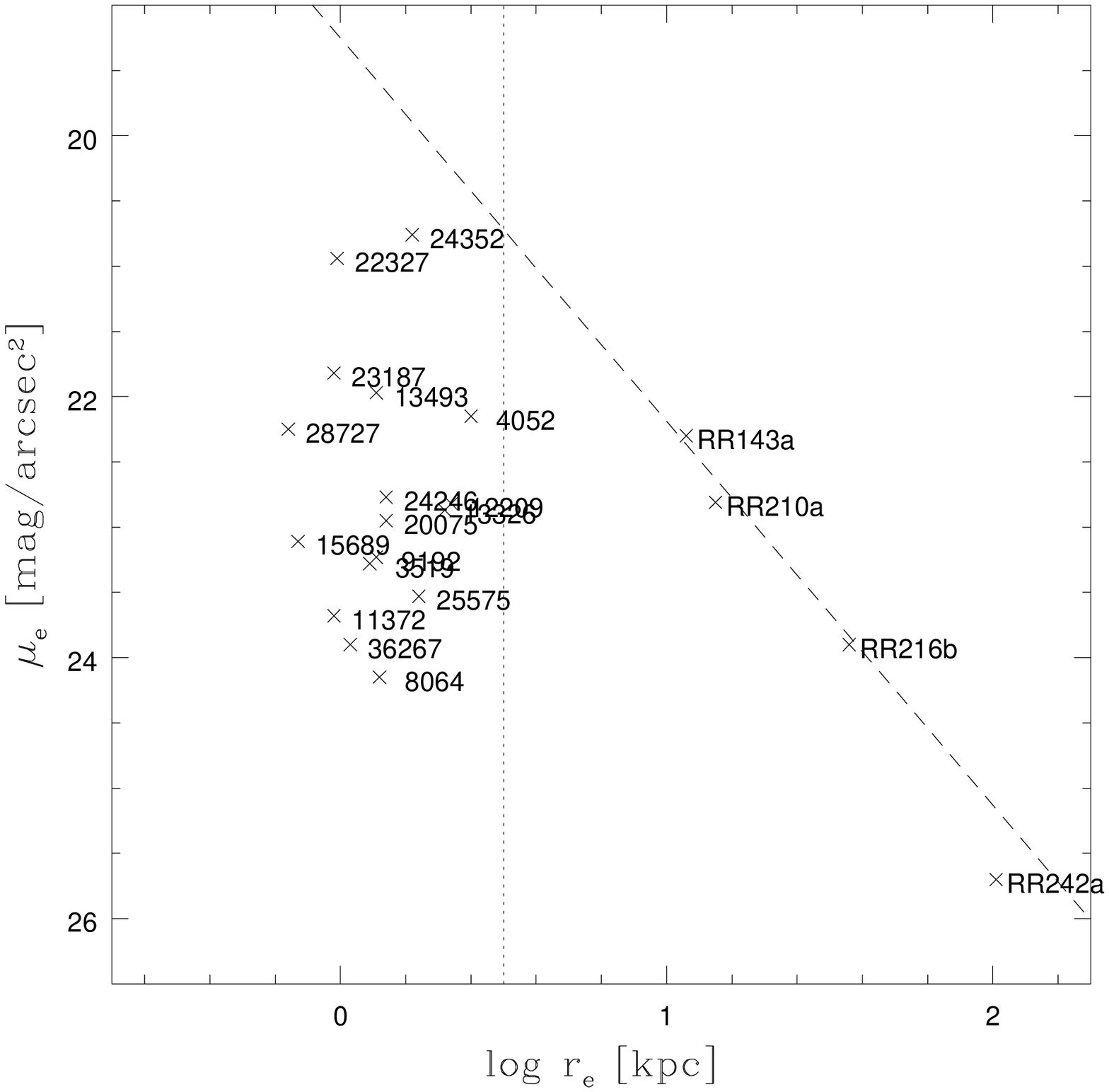}
\caption{Hamabe-Kormendy relation \citep[][dashed line]{HK87} for new group member galaxies. The pair ellipticals are also plotted for comparison. All galaxies are labelled with their ID. The area left of the vertical dotted line represents the region inhabited by ``ordinary'' galaxies \citep[see][]{cap92}.
\label{fig11}}
\end{figure}

In order to investigate the bulge-to-total light (B/T) ratios a bulge-disk decomposition was attempted with {\tt GALFIT} \citep{Peng02}. \citet{Tran01}
found that the galaxy population of poor X-ray detected groups was well described by a 2-component model composed of a de Vaucouleurs bulge ($\mu_r \sim
r^{1/4}$) and an exponential disk ($\mu_r \sim r$). However, the shape of the profile determined by the exponent $1/n$ is supposed to vary significantly with mass. This is valid not only for low-mass ellipticals but also for the bulges of low-mass galaxies, which are expected to show a different profile shape, i.e., a lower $n$. Both bulge/disk combinations with $n=4$ (de Vaucouleurs bulge) and $n$ as a free parameter (Sersic bulge) were fit to our member galaxies to investigate differences between these two models. Bulge and disk magnitudes as well as B/T ratios for the two different models can be found in Table~\ref{table4}. Differences in B/T can be significant especially for bulge-dominated galaxies. The difference between the $\chi^2_\nu$ of the two-component fits from {\tt GALFIT} is given in the last column of Table~\ref{table4}. We find that the Sersic bulge provides an accurate representation of faint galaxy bulges: apart from one galaxy (RR242\_28727), the Sersic fit always has a lower $\chi^2_\nu$ than the de Vaucouleurs fit. The de Vaucouleurs fits of RR143\_09192, RR210\_13493, RR242\_24352 and RR242\_36267 are unsatisfactory, while RR242\_23187 was fitted badly by the Sersic model.

Figures~\ref{fig8} and \ref{fig9} show the results of the Sersic-model fit and the observed surface-brightness profile in the R-band. The dotted, red lines represent the bulge and disk model, respectively. A model for the bar was added for the brighter bars. The solid line represents the resulting galaxy model. 

Figure~\ref{fig10} (top panel) shows the distribution of B/T ratios for Sersic (black, shaded) and de Vaucouleurs models (red, dotted line). Both distributions show that our sample is dominated by low B/T - i.e., disk-dominated - galaxies. An automatic classification by the B/T ratio was proposed by \citet{Mar98} and also used by \citet{Tran01}. This automatic classification divides late (S, disk-dominated) and early-type (S0 and E, bulge-dominated) galaxies at B/T = 0.4. This may work in distinguishing between bright E and S galaxies, but is problematic for faint S0 galaxies. A comparison between visual (based on surface photometric profiles) and automatic classification (based on the B/T ratio) shows the problem: visually classified S0 galaxies show a wide range of B/T ratios and are not necessarily bulge-dominated systems. We found a high fraction of S0s from our visual classification (7 out of 17, 40\%), but only 3 of those galaxies have a B/T $\geq$ 0.4 (18\%). This would yield an early-type fraction similar to that of the field, while the visually
estimated S0-fraction is more typical of galaxy clusters and X-ray luminous galaxy groups \citep[][and references therein]{Tran01}.

Figure~\ref{fig10} (middle panel) plots B/T ratios versus the projected distance from the dominant E member. The individual galaxies are plotted as small dots, while the mean and dispersion are computed in bins of 30 $h_{100}^{-1}$ kpc and plotted as open circles with respective error bars (red -- de Vaucouleurs bulge; black -- Sersic bulge). The lines are a least squares fit to the individual data points (solid black line -- Sersic; dashed, red line -- de Vaucouleurs). A morphology-radius relation appears to be present in our sample: galaxies with higher B/T ratios tend to be more centrally concentrated than pure disks. This result was also found for the groups studied by \citet{Tran01}. However, the relation is very flat and the scatter is large, probably due to projection effects and the small number of galaxies studied. We computed Spearman rank coefficients for the two data sets resulting in $\rho_{sersic} = 0.14$ and $\rho_{deV} = 0.32$, values that are not considered to be statistically significant. 

\begin{figure*}
\hspace{-0.5cm}
\includegraphics[angle=270, width=18cm]{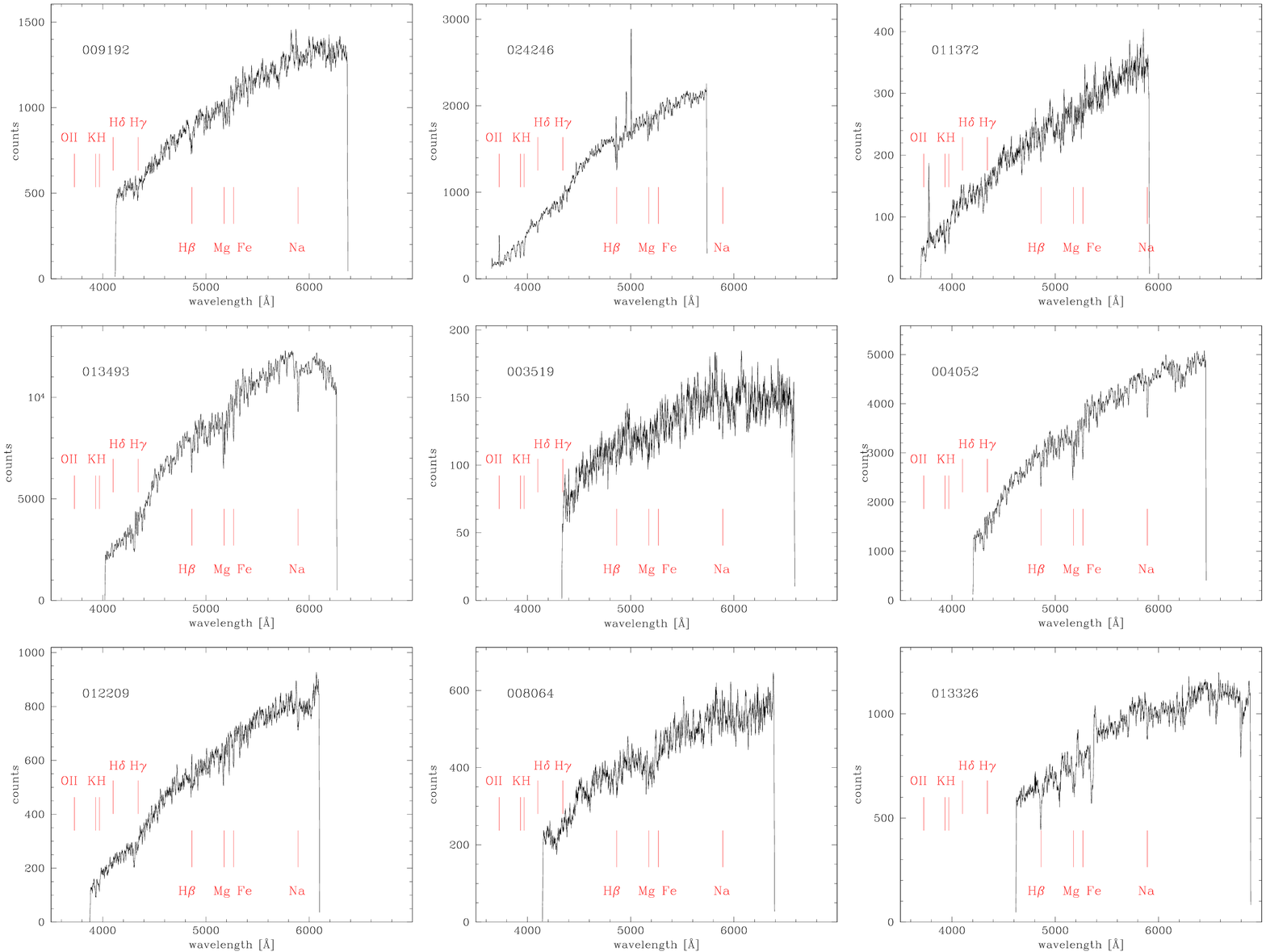}
\caption{Rest frame spectra of member galaxies observed with VIMOS. Objects RR143\_09192, RR143\_24246, RR210\_11372, RR21\_13493, RR216\_03519,
RR216\_04052, RR216\_12209, RR242\_08064, and RR242\_13326. The most prominent absorption and emission lines are marked.
\label{fig12}}
\end{figure*}

The bottom panel of Fig.~\ref{fig10} shows the dependence of B/T on the local projected number density. This local density was computed using the distance to the 5$^{th}$ ($d_5$) and 3$^{rd}$ ($d_3$) closest group member. The results are plotted as circles ($d_5$) and triangles ($d_3$), and in black and red for Sersic and de Vaucouleurs bulges, respectively. The lines are the least squares fit to the data of $d_5$ (dashed-dotted -- Sersic; red dashed -- de Vaucouleurs) and $d_3$ (long-dashed -- Sersic; red short-dashed -- de Vaucouleurs). Using only the area occupied by the 3 closest galaxies changes the results slightly but significantly: the correlation for $d_5$ is stronger than for $d_3$, the Spearman rank coefficients being $\rho_{d_3} = 0.46$ and $\rho_{d_5} = 0.56$. For our sample size (16 galaxies), the latter value is higher than the critical value of $\rho$ at the 0.05 (2$\sigma$) level of significance. Hence, there seems to be a positive correlation between the projected number density ($d_5$) and the morphology (expressed by the B/T ratio).\\

The Hamabe-Kormendy Relation \citep[][HKR hereafter]{HK87} in the $\log r_e - \mu_e$ plane is one projection of the Fundamental Plane for early-type galaxies. Faint Es, S0s, or dwarf ellipticals (dEs) do not follow this relation but are distributed in the $r_e - \mu_e$ plane below the HKR and also below $r_e = 3$ kpc. They are considered a distinct family of galaxies, the so-called {\it ordinary} galaxies, whereas galaxies above $r_e = 3$ kpc belong to the family of {\it bright} ellipticals \citep{cap92}. The ordinary family are often considered to be the building blocks of galaxies of the bright family, although more recent simulations of \citet{evs04} show that by merging, the galaxies evolve along tracks that are parallel to the HKR.

The $r_e$ - $\mu_e$ plane for our four E+S groups is shown in Fig.~\ref{fig11} with the dashed line indicating the HKR (in the R-band). The dotted line at $\log r_e = 0.5$ separates the bright and ordinary families \citep{cap92}. The four bright Es lie on the HKR clearly in the bright galaxy domain, while the newly identified faint members are distributed in the $\log r_e - \mu_e$ plane below the limit of ordinary galaxies. The brightest galaxy of the newly discovered galaxy population RR242\_24352 is closest to the HKR, while the faintest galaxies are off the relation in the expected vertical strip of ordinary galaxies \citep{cap92}.

\begin{figure*}
\hspace{-0.5cm}
\includegraphics[angle=270, width=18cm]{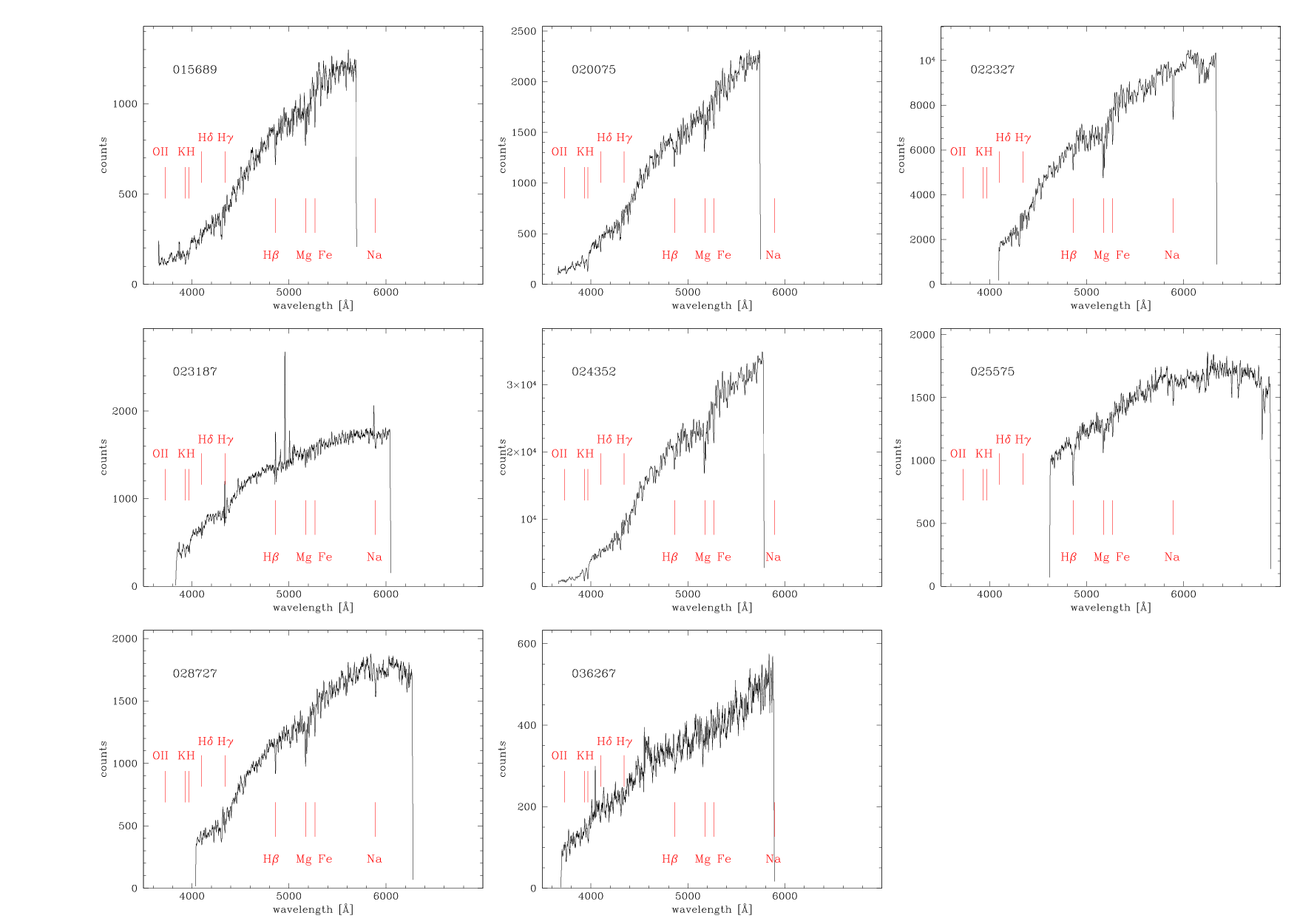}
\caption{Rest frame Spectra of member galaxies observed with VIMOS.  Objects
RR242\_15689, RR242\_20075, RR242\_22327, RR242\_23187, RR242\_24352,
RR242\_25575, RR242\_28727, and RR242\_36267. The most prominent absorption
and emission lines are marked.
\label{fig13}}
\end{figure*}

\subsection{The spectral properties of faint members}

VIMOS spectra of the new member galaxies are presented in Figs.~\ref{fig12} and \ref{fig13}. As already suggested by the colour-magnitude relation, the
spectra are characterised by a relatively old stellar population.  Strong metal lines (Mg I and Fe) are present in most of the spectra, although many also exhibit strong H$\beta$ absorption suggesting the presence of a younger or intermediate-age population.  Emission lines are detected in only two galaxies; RR242\_23187, where H$\gamma$ and H$\delta$ emission suggests recent or ongoing star-formation activity, and RR143\_24246, where [O~II] $\lambda$ 3727-29 \AA\ emission is detected.  The forbidden lines of [O~III] $\lambda$ 4959 \AA\ and $\lambda$ 5007 \AA\  as well as H$\beta$ emission is present in both galaxies, although on top of a substantial absorption component. The detailed analysis of the line-strength indices \citep[see e.g.,][]{Rampazzo05,Gru05,Annibali07} will be treated in a forthcoming paper.

\section{Discussion}\label{discussion}

In the following discussion, we consider only the spectroscopically confirmed member galaxies. Their basic properties are summarised in Table~\ref{table3}. The new members are used, along with the previously known group members (see Paper~III), to infer group properties. Table~\ref{tableA1} lists all known members of the groups: i.e., new members found within the WFI field of view and already known members found in the NASA/IPAC Extragalactic Database (NED). To investigate the effects of the WFI's small field of view, we compare velocity and magnitude distributions of the full sample (90\arcmin) and the WFI subsample (34\arcmin$\times$34\arcmin).

Group kinematics and dynamics are discussed in the first section using a luminosity-weighted approach for the determination of all mass-related quantities \citep[see e.g.,][]{Firth06}. However, we also compute uniformly weighted quantities and discuss the results of the different weighting. The dynamical formulae are given in the Appendix. Distribution of members, group compactness, crossing times, and mass-to-light ratios are analysed and compared with the literature. OLFs of the individual and combined group samples are presented in the next section. The OLFs of our X-ray faint and X-ray bright groups are compared with OLFs for samples of: 1) X-ray detected poor groups, 2) simulated and observed fossil groups, and 3) the OLF of the local field. We then attempt to investigate the possible relation between the group dynamical characteristics as well as the group ``activity'' and the X-ray luminosity of the E members.

\subsection{The E+S system kinematics and dynamics}

\begin{figure}
\includegraphics[width=9cm]{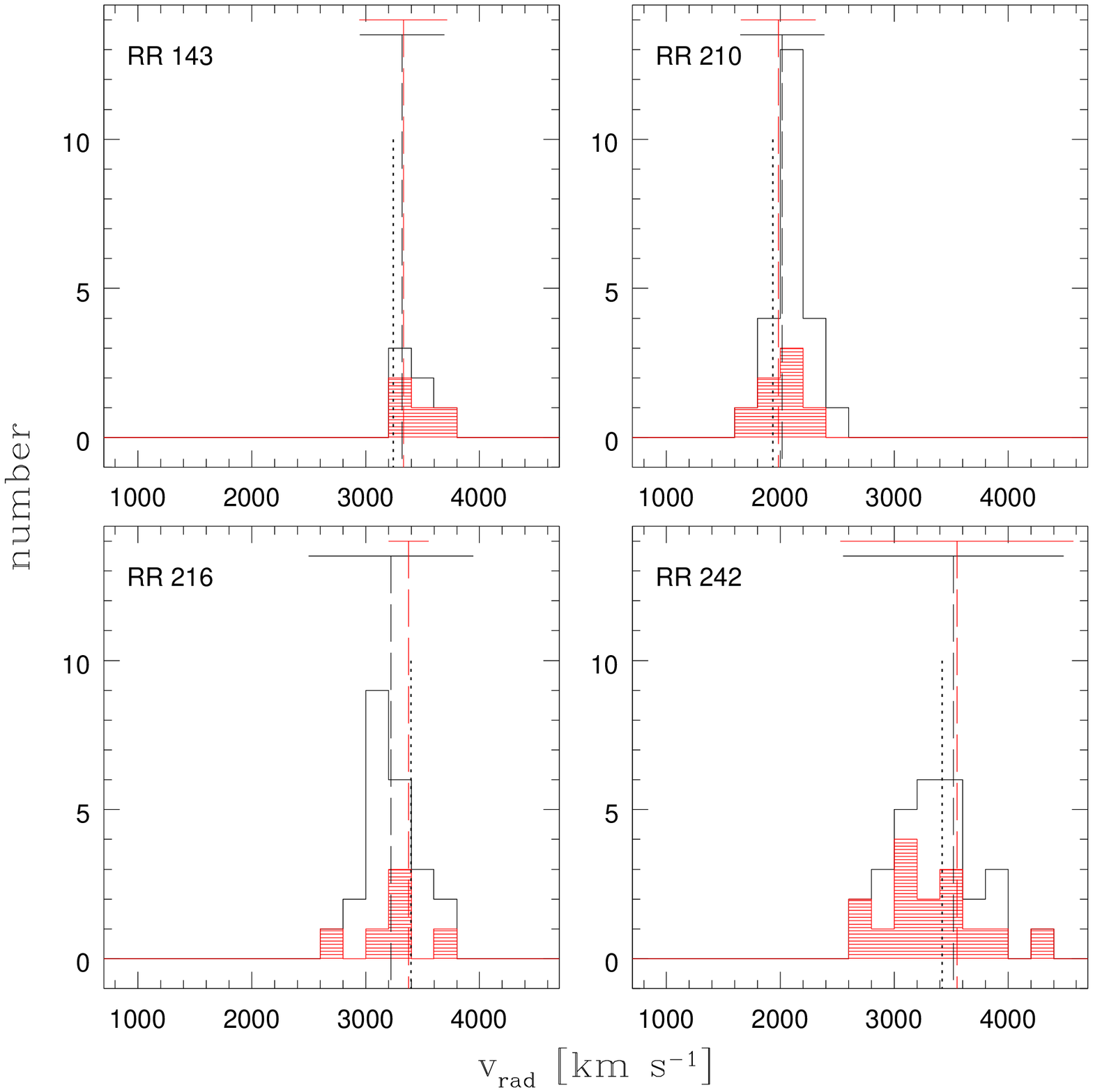}
\caption{Distribution of radial velocities in the 4 groups, including previously known and new member galaxies. The luminosity-weighted mean velocity and 3$\sigma$ velocity dispersion are plotted as vertical dashed and horizontal solid line respectively. The WFI-subsample is plotted in red (dashed), the line indicating the 3$\sigma$ velocity dispersion of the WFI-subsample is drawn above the black line indicating the 3$\sigma$ velocity dispersion of the whole sample within 90$^\prime$ from the pair E. The velocity of the elliptical pair member is indicated by the shorter dotted line.
\label{fig14}}
\end{figure}

Distributions of radial velocities for the full sample and the WFI-sub\-sample (red-dashed) are shown in Fig.~\ref{fig14}. The $v_{group}$ of each sample is plotted as a vertical, dashed line, while the horizontal line above each histogram indicates the 3 $\sigma_r$ limits as an approximate dynamical boundary for each group. The mean velocities of the two samples do not differ significantly, apart from RR~216, where the velocity of the WFI-subsample is dominated by the very bright pair elliptical (no other bright members are present in the WFI-field) and is therefore biased towards a higher value. For this group, the group velocity for the full sample is $v_{90arc} = 3223\pm49$ versus $v_{WFI} = 3378\pm24$ for galaxies in the WFI field of view. The unweighted velocity dispersions of the two samples are comparable within the errors for all four systems. The luminosity-weighting significantly changes the velocity dispersion only in RR~216 where the velocity difference between the 2 dominating pair galaxies is very small, biasing the dispersion within the WFI field towards a low value ($\sigma_{90arc} = 241\pm35$ versus $\sigma_{WFI} = 56\pm18$). \\

\begin{figure}
\includegraphics[width=9cm]{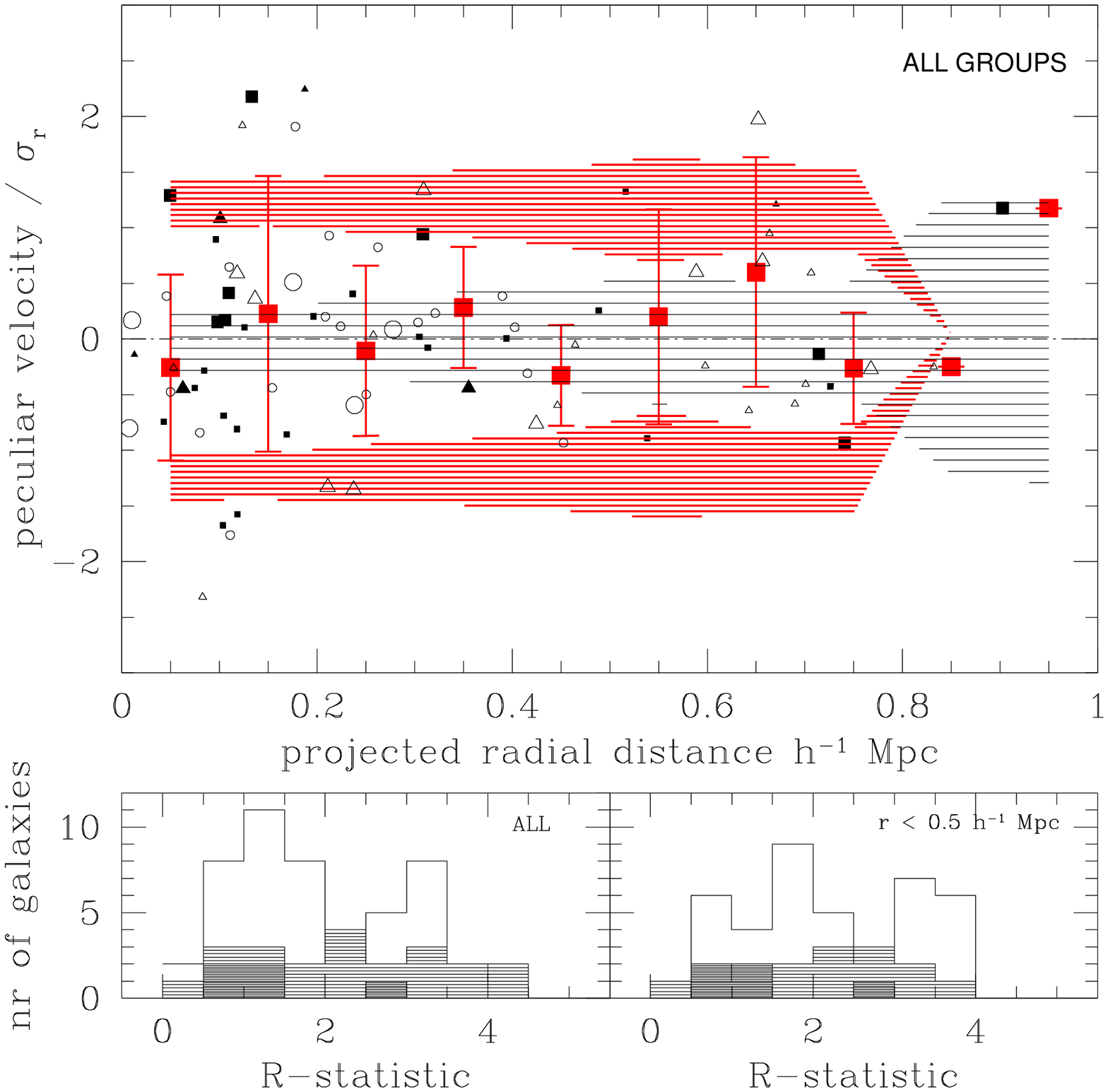}
\caption{Analysis of galaxy offsets in projection and velocity. {\it Upper panel:} Peculiar galaxy velocities ($v_{galaxy} - v_{group}$) normalised by the line-of-sight velocity dispersion $\sigma_r$ as a function of the projected radial distance from the optical group centre. Members of all 4 group are plotted  with different symbols: solid triangles: RR~143; open circles: RR~210; open triangles: RR~216; solid squares: RR~242. The size of the symbols indicate dwarf (small symbols) and giant (big symbols) group members. The galaxies are binned in projected radial distance every 0.1 $h_{100}^{-1}$ Mpc. The mean velocity and velocity dispersion in each bin is plotted as red squares and red error bars. The shaded areas show the $1 \sigma$ deviations expected from the Monte Carlo simulation (see text). The mean velocity rms is mildly shaded (in grey), while the area of the $1 \sigma$ deviation from a constant velocity dispersion is heavily shaded (in red). {\it Bottom panels:} The $R$-statistic quantifies the offset in velocity and distance of each galaxy from the group mean velocity and from the optical group centre (see text). BGGs (heavily shaded), giants (shaded), and dwarfs (unshaded) show different distributions of $R$ suggesting that they occupy different orbits.
\label{fig15}}
\end{figure}

The peculiar velocity is the difference between an individual galaxy velocity and the group centre velocity ($v_{galaxy} - v_{group}$). It is usually normalised by the group's velocity dispersion for comparison with other groups \citep[see e.g.,][]{ZM00}. Figure~\ref{fig15} shows the peculiar velocities of galaxies versus their projected radial distance from the optical group centre. The group members are separated into {\it giants} (big symbols) and {\it dwarfs} (small symbols) following \citet{ZM00}. Galaxies with absolute magnitudes brighter than $M_R = -19 +5 \log h_{100}$ are considered giants, and fainter objects are defined as dwarfs. The four group samples are combined in  analysing the dependence of velocity dispersion on the projected distance. The mean velocity and velocity dispersion is computed in radial bins of 100 $h_{100}^{-1}$ kpc. They are plotted as red squares (mean velocity) and respective red error bars (velocity dispersion) for each bin.

Figure~\ref{fig15} suggests that the velocity dispersion is not constant with projected radius. The maximal dispersion is reached at around 0.2 $h_{100}^{-1}$ Mpc ($\sim$ the border of the WFI field of view), from where it starts to decrease out to $\sim$ 0.5 $h_{100}^{-1}$ Mpc. At greater radii, the dispersion rises again, which could indicate the transition between the potential of the group concentrated around the E and the influence of the global large-scale density. We computed the statistical errors in the velocity dispersion in each bin following \citet[][their equation 4]{Osm04}. The $\sigma_{min}$ and $\sigma_{max}$ quoted below are normalised velocity dispersions obtained by dividing the peculiar galaxy velocities by the respective group's velocity dispersion. The maximum of $\sigma_v$ at 0.15 $h_{100}^{-1}$ Mpc is $\sigma_{max} = 1.24\pm0.20$, while the minimum at 0.45 $h_{100}^{-1}$ Mpc is $\sigma_{min} = 0.45\pm0.14$. Hence, the velocity dispersion is not constant within the errors. However, this is a tentative result due to the intrinsically low number of members in our groups.

To investigate the effect of incompleteness and low number statistics in our sample, a set of Monte Carlo simulations was performed. The question is whether the detected drop in velocity dispersion is significant, or if our measured $\sigma$ cannot be distinguished from a constant velocity dispersion. Therefore, in each radial bin a Gaussian velocity distribution of the same $\sigma$ (the maximum velocity dispersion found in the second radial bin: $\sigma_{max} = 1.24$) was assumed. Then, a random sample of $n$ velocities was taken from this Gaussian distribution, where $n$ represents the number of galaxies of our sample in the respective bin. After 1000 iterations, the mean velocity and velocity dispersion and their deviations in each bin was computed. The $1 \sigma$ deviations are plotted in Fig.~\ref{fig15} as grey (mean velocity) and red (velocity dispersion) shaded areas. The result is that the mean velocity is consistent with the group velocity in all bins, whereas the velocity dispersion is lower than the expected $1 \sigma$ deviation from the constant value in all bins out to 0.5 $h_{100}^{-1}$ Mpc. If the velocity dispersion were constant, the error bars would reach into the red shaded area. 
This result is significant at $> 2 \sigma$ in the bin between 0.4 - 0.5 $h_{100}^{-1}$ Mpc. We repeated this analysis with different radial binnings (75, 100, and 150 $h_{100}^{-1}$ kpc) and the drop in velocity dispersion was always above the 1$\sigma$ significance level. 
We therefore suggest that the velocity dispersion decreases with radius and that the kinematics in the region outside 0.5 $h_{100}^{-1}$ Mpc may not trace the group potential (as traced by the IGM). The dynamical quantities were also calculated by excluding galaxies lying outside this radius.

\begin{table*}
\begin{scriptsize}
\begin{center}
\caption{Dynamical properties of the E+S systems. }
\label{table5}
\begin{tabular}{lccccccc}
\hline\hline
Group & Nr. of & Distance & \multicolumn{2}{c}{Optical group centre} & $v_{group}$ & Velocity & 3D velocity \\
     & members & (Modulus) & $\alpha$ (2000)& $\delta$ (2000)& & dispersion & dispersion \\
     & & [Mpc (mag)] & [h:m:s] & [$^\circ$:$^\prime$:$^{\prime\prime}$] & [km s$^{-1}$] & [km s$^{-1}$] & [km s$^{-1}$] \\
\hline
 & & & & & & & \\
RR~143        &  6 & 49.6 (33.5) & 06:48:05.6 & -64:10:54 & 3321$\pm$52 & 126$\pm$39 & 214 \\
              &  5 & 49.5 (33.5) & 06:47:59.2 & -64:11:57 & 3317$\pm$53 & 123$\pm$42 & 210 \\
Uniform $w_i$ &  6 & 51.1 (33.5) & 06:48:28.9 & -64:02:09 & 3426$\pm$72 & 180$\pm$54 & 304 \\
              &  5 & 50.8 (33.5) & 06:47:22.7 & -64:11:57 & 3404$\pm$82 & 189$\pm$66 & 319 \\
    & & & & & & & \\
RR~210        & 23 & 30.1 (32.4) & 12:06:55.8 & -29:46:47 & 2018$\pm$26 & 123$\pm$18 & 210 \\
Uniform $w_i$ & 23 & 31.5 (32.5) & 12:06:42.8 & -29:44:27 & 2113$\pm$34 & 165$\pm$24 & 273 \\
    & & & & & & & \\
RR~216        & 23 & 48.1 (33.4) & 12:26:05.5 & -39:38:38 & 3223$\pm$49 & 241$\pm$35 & 414 \\
              & 12 & 47.6 (33.4) & 12:26:01.5 & -39:33:50 & 3191$\pm$67 & 229$\pm$48 & 393 \\
Uniform $w_i$ & 23 & 48.6 (33.4) & 12:26:17.1 & -39:40:19 & 3255$\pm$52 & 241$\pm$36 & 411 \\
              & 12 & 47.9 (33.4) & 12:25:56.4 & -39:37:55 & 3206$\pm$80 & 273$\pm$58 & 463 \\
    & & & & & & & \\
RR~242        & 28 & 52.5 (33.6) & 13:20:20.7 & -43:42:06 & 3520$\pm$60 & 323$\pm$45 & 537 \\
              & 22 & 52.9 (33.6) & 13:20:54.8 & -43:42:08 & 3546$\pm$65 & 314$\pm$48 & 529 \\
Uniform $w_i$ & 28 & 50.1 (33.5) & 13:20:21.6 & -43:38:07 & 3356$\pm$70 & 368$\pm$51 & 623 \\
              & 22 & 50.1 (33.5) & 13:20:44.1 & -43:37:00 & 3354$\pm$76 & 363$\pm$56 & 617 \\
 & & & & & & & \\
Group & Harmonic & Virial & Crossing &Virial & Projected & Group & $M_\odot/L_\odot$ \\
     & radius ($R_H$) & radius ($R_{vir}$) & time & mass & mass & luminosity & \\
     & [Mpc] & [Mpc] & [$t_c$ H$_0$] & [10$^{12}$ M$_\odot$] & [10$^{12}$ M$_\odot$] & [10$^{11}$ L$_\odot$] & \\
& & & & & & & \\
RR~143        & 0.170$\pm$0.001 & 0.267$\pm$0.001 & 0.14$\pm$0.02 & 1.5$\pm$0.7 & ~7.8$\pm$6.5  & 1.06$\pm$0.005 &  13$\pm$1 \\
              & 0.161$\pm$0.001 & 0.252$\pm$0.001 & 0.09$\pm$0.02 & 1.3$\pm$0.7 & ~5.7$\pm$4.7  & 1.04$\pm$0.004 &  12$\pm$1 \\
Uniform $w_i$ &  0.43$\pm$0.04  &  0.68$\pm$0.06  & 0.16$\pm$0.09 & 7.6$\pm$5.0 & 30.2$\pm$20.6 & 1.13$\pm$0.005 &  66$\pm$5 \\
              &  0.31$\pm$0.03  &  0.49$\pm$0.05  & 0.09$\pm$0.07 & 6.1$\pm$4.8 & 21.8$\pm$10.9 & 1.10$\pm$0.005 &  55$\pm$5  \\
 & & & & & & & \\
RR~210        & 0.053$\pm$0.0001 & 0.083$\pm$0.0002 & 0.14$\pm$0.01 & 0.4$\pm$0.2 & ~6.0$\pm$4.2  & 1.56$\pm$0.001 & ~~2$\pm$0.2 \\
Uniform $w_i$ &  0.44$\pm$0.06   &  0.70$\pm$0.10   & 0.17$\pm$0.21 & 6.5$\pm$4.3 & 16.6$\pm$12.0 & 1.71$\pm$0.001 & 38$\pm$4 \\
 & & & & & & & \\
RR~216        & 0.35$\pm$0.01 & 0.55$\pm$0.01 & 0.24$\pm$0.03 & 11.1$\pm$3.3  & 63.2$\pm$39.9 & 3.84$\pm$0.010 & 28$\pm$3 \\
              & 0.24$\pm$0.01 & 0.37$\pm$0.01 & 0.10$\pm$0.02 & ~6.7$\pm$2.8  & 32.0$\pm$22.0 & 3.14$\pm$0.008 & 21$\pm$3 \\
Uniform $w_i$ & 0.87$\pm$0.09 & 1.36$\pm$0.14 & 0.24$\pm$0.19 & 27.3$\pm$10.0 & 65.4$\pm$42.8 & 3.92$\pm$0.010 & 69$\pm$10 \\
              & 0.45$\pm$0.06 & 0.71$\pm$0.10 & 0.10$\pm$0.09 & 18.3$\pm$9.4  & 37.5$\pm$20.4 & 3.71$\pm$0.009 & 57$\pm$9 \\
 & & & & & & & \\
RR~242        & 0.221$\pm$0.001 & 0.346$\pm$0.001 & 0.09$\pm$0.01 & 12.5$\pm$2.8 & ~92.3$\pm$85.0 & 2.19$\pm$0.005 & 57$\pm$3 \\
              & 0.141$\pm$0.001 & 0.221$\pm$0.001 & 0.05$\pm$0.01 & ~7.6$\pm$1.8 & ~31.6$\pm$17.7 & 1.79$\pm$0.005 & 42$\pm$2 \\
Uniform $w_i$ &  0.41$\pm$0.02  &  0.64$\pm$0.02  & 0.10$\pm$0.03 & 29.9$\pm$9.1 & 119.0$\pm$97.3& 1.99$\pm$0.004 & 150$\pm$9 \\
              &  0.28$\pm$0.01  &  0.44$\pm$0.02  & 0.06$\pm$0.02 & 20.3$\pm$6.9 & ~59.3$\pm$23.8& 1.61$\pm$0.004 & 125$\pm$7 \\
\hline
\multicolumn{8}{l}{}\\
\multicolumn{8}{l}{Notes: The first two rows give the luminosity weighted results for the whole sample (1$^{st}$ row) and for galaxies within} \\
\multicolumn{8}{l}{0.5 $h_{100}^{-1}$ Mpc (2$^{nd}$ row). In line 3 and 4 the same values are calculated with uniform weights.} \\
\end{tabular}
\end{center}
\end{scriptsize}
\end{table*}

Another  interesting implication of Fig.~\ref{fig15} is that the position of the bright E+S pair does not coincide with the optical centre of each group. The projected offsets for RR~216, RR~242, and RR~143 are around 0.1 $h_{100}^{-1}$ Mpc, and only RR~210 is located very close to the projected group centre. 

The $R$-statistic was developed by \citet{ZM00} to facilitate comparison between the distribution of members in both projection and velocity space. It is defined as $R^2 = (d/\delta_d)^2 + (|v_{pec}|/\delta_{|v_{pec}|})^2$, where $\delta_d$ and $\delta_{|v_{pec}|}$ denotes the rms deviations of the entire sample in projected distance and peculiar velocity, respectively. A galaxy with a large distance $d$ from the group centre or a large peculiar velocity $v_{pec}$ will yield a high value of $R$, while an average member should have $R \sim 2$. We compared the distribution of $R$ values for three magnitude-limited subsamples: 1) the brightest group galaxies (BGGs) with $M_R < M^\ast$, 2) giants with $M_R < -19 +5 \log h_{100}$, and 3) dwarfs. We also compared the full sample with galaxies within 0.5 $h_{100}^{-1}$ Mpc of the group centre.  Figure~\ref{fig15} (bottom panels) shows the distribution of $R$ for the BGGs (heavily shaded), giants (shaded), and dwarfs (unshaded) in the full sample (left panel) and for the galaxies within 0.5 $h_{100}^{-1}$ Mpc (right panel). In both samples, the BGGs are more centrally concentrated than dwarfs and giants with $\langle R_{BGG} \rangle = 1.3\pm0.7$ and $\langle R_{BGG,0.5} \rangle = 1.4\pm0.8$.  Considering the full sample, the dwarf population follows a different distribution in phase-space than the giants with $\langle R_{dwarfs} \rangle = 2.0\pm0.9$ and $\langle R_{giants} \rangle = 2.3\pm1.1$.  

A K-S test is used to check whether the $R$ distributions of the three subsamples differ significantly. It gives the following probabilities that the 3 distributions are the same: $P_{BGG,DWARF} = 0.13$, $P_{BGG,GIANT} = 0.23$, and $P_{DWARF,GIANT} = 0.29$. Those values are significant above the 1$\sigma$ level. Inside the 0.5 $h_{100}^{-1}$ Mpc radius, the difference between dwarfs and giants vanishes (both have $\langle R \rangle \sim 2.2$), but the $R$ distributions of BGGs and dwarfs have a probability of originating in the same distribution of $P_{BGG,DWARF} = 0.04$. This suggests that the 2 distributions are different above the 2 $\sigma$ significance level. This is not only caused by the optical group centre moving closer to the BGGs, since both, $\langle R_{BGG,0.5} \rangle$ and $\langle R_{DWARF,0.5} \rangle$ are higher than the respective values for the full sample.

These findings are consistent with the results of \citet{ZM00}, who argued that the three galaxy populations (BGGs, giants, dwarfs) move on different orbits and have not yet mixed. From the present data, this also seems to be the case for the BGGs and dwarfs at the centres of our 4 groups, although we note that outside the WFI field of view we lack information about faint member galaxies.\\

The dynamical properties of the four E+S systems are given in Table~\ref{table5}, where the first row provides data for the entire sample and the second row for group members inside a radius of 0.5 $h_{100}^{-1}$ Mpc. Errors were estimated using Monte Carlo simulations. For velocity-related quantities ($v_{group}$, velocity dispersion), a Gaussian distribution was assumed. A set of 1000 groups with the respective number of members was constructed for both the full sample and galaxies within 0.5 $h_{100}^{-1}$ Mpc (see Appendix~A). Each of these groups consisted of galaxies with velocities that were taken randomly from a Gaussian distribution of respective mean velocity and dispersion given in Table~\ref{table5}. The rms of the mean velocity and velocity dispersion of this set of groups is the error given in Table~\ref{table5}. The position and luminosity related quantities were treated in a different way. Here, the number of additional galaxies expected from the completeness correction was added to the existing group. These additional galaxies were selected randomly from the sample of candidates without a measured redshift. After 1000 iterations, the rms of the values ($R_H$, $R_{vir}$, $r_\perp$, group luminosity) of this set of ``complete'' groups were computed. The errors in virial mass $M_{vir}$, projected mass $M_P$, and crossing time $t_c$ are a combination of this two methods, since they are dependent on velocity and position of the objects.

\begin{figure}
\includegraphics[width=9.2cm]{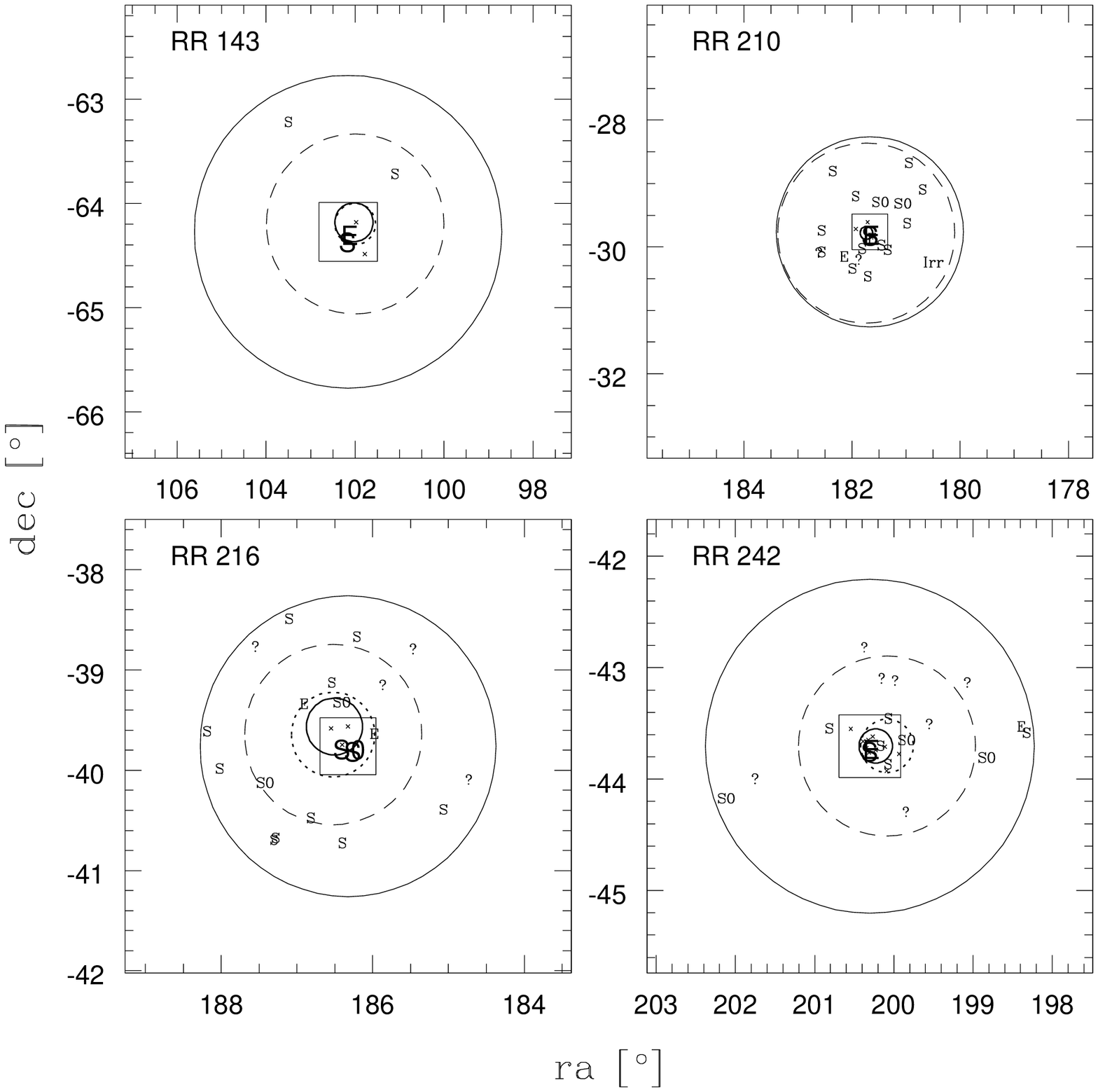}
\caption{Position of member galaxies of the 4 groups moved to a common distance. The field of view of each plot is $\sim$ 2.5 $h_{100}^{-1}$ Mpc on each
side. The positions of all galaxies with concordant radial velocity found in NED within 90$^\prime$ are indicated with each galaxy's morphological type. The 90$^\prime$ search radius around the E pair member is indicated by the solid circle. New members found in the WFI field (central square) are plotted as crosses. The dashed circle represents the 0.5 $h_{100}^{-1}$ Mpc limit found in Fig.~\ref{fig15}, beyond which the velocity dispersion starts to rise again.  The central circle is the luminosity-weighted mean harmonic radius $R_H$ centred on the optical group centre calculated with group members inside the 0.5 $h_{100}^{-1}$ Mpc radius (solid line) and with the whole sample (dotted line), see Table~\ref{tableA1} and Table~\ref{table5} for the values.
\label{fig16}}
\end{figure}

Both luminosity-weighted and uniformly-weighted results are provided in Table~\ref{table5}. Different weightings influence the results dramatically, and especially in RR~210 where the parameters of the bright pair dominate the resulting values. The pair components are separated by a projected distance of only 6 $h_{100}^{-1}$ kpc and are close to the centres of both mass and velocity. This leads to an underestimation of $M_{vir}$ and a very low luminosity-weighted value of $M/L$ for this pair. $R_H$ and $M_{vir}$ rise by an order of magnitude if the members are uniformly weighted. Values change by a factor of $\sim$ 2 for the other systems. The group velocity and velocity dispersion are less affected by the weighting. Weights given in Table~\ref{tableA1} illustrate the dominance of the E member. RR~143a and RR~210a contribute more than 50\% to the total luminous mass, while the spiral companion is clearly the second brightest object. RR~216b and RR~242a contain about 1/3 of the luminous mass and have a few other massive objects brighter or comparable to the spiral pair member in their close environment. The unweighted value of $R_H$ for RR~216 is very high and indicative of a higher large-scale density in which this pair is embedded and a probable dynamical link to the Hydra-Centaurus cluster. However, it also means that the brighter members are more
concentrated towards the pair than the intermediate-luminosity galaxies.

\begin{figure}
\includegraphics[width=9cm]{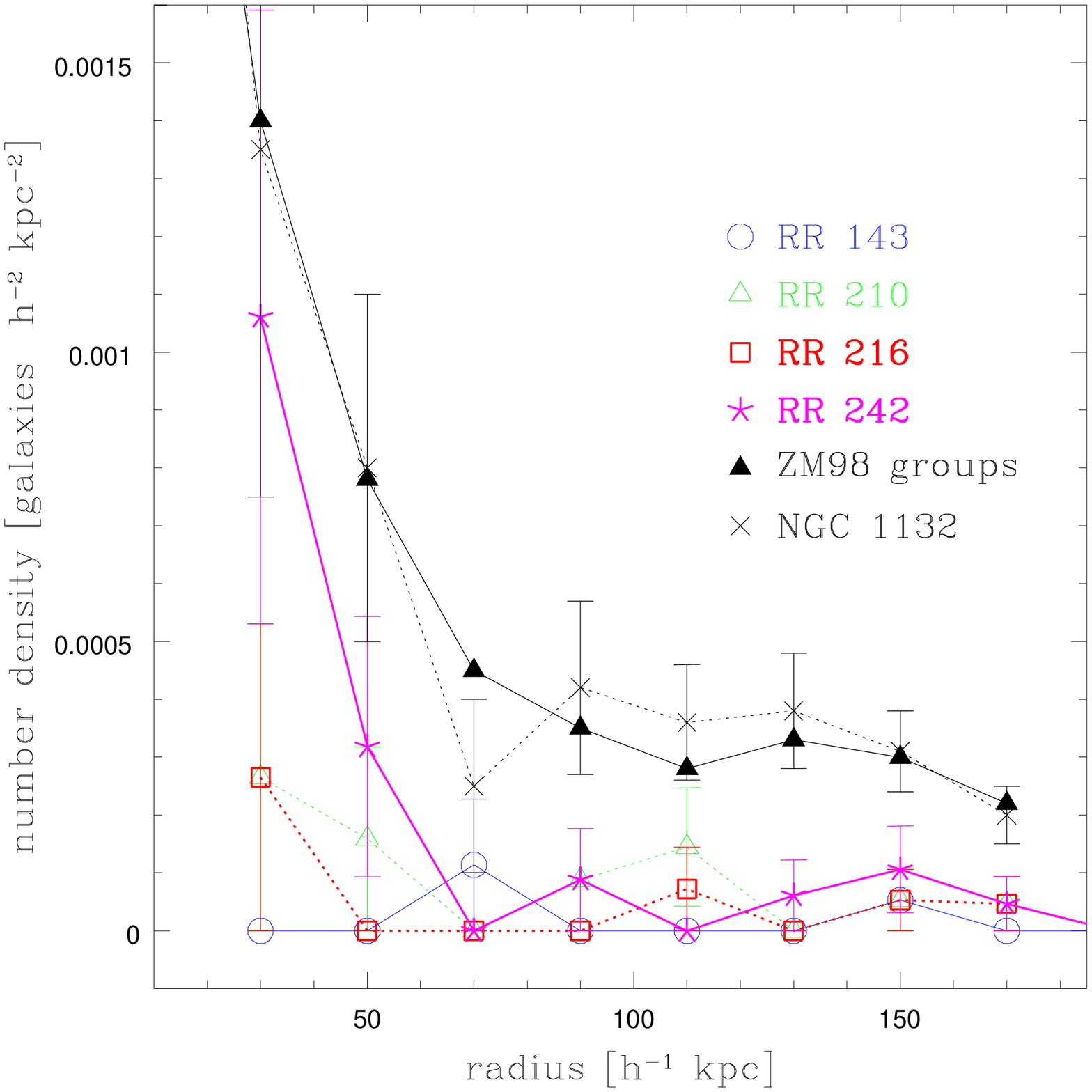}
\caption{Radial density profile of faint group member galaxies. The groups are colour-coded and marked with different symbols. Note the concentration of
faint galaxies towards the centre (the pair E member) in RR~242, but not in RR~143. For comparison are shown the radial distribution of galaxies in the groups studied by \citet{ZM98} and of the galaxies in the fossil group around NGC~1132 \citep[see][]{Mul99}.
\label{fig17}}
\end{figure}

Differences between $M_{vir}$ and $M_P$ are quite high with $M_P$ being generally higher than $M_{vir}$ by a factor of 2-5. This effect is expected in systems where individual galaxies are close to each other in projection. $M_{vir}$ was found to underestimate the mass by a factor of $\sim$ 3 or more in such systems \citep{Heisler85}. An alternative reason for this difference could be that the groups are not virialised. On the other hand, the crossing times of the groups are a small fraction of a Hubble-time and usually shorter than 0.2 (apart from RR~216), even when calculated with the entire sample. This indicates that the groups, or at least their centres, are virialised \citep{Fergusson90}. RR~216 and RR~242 have comparable $M_{vir}$ but very different $M/L$ due to the very high group luminosity of RR~216, which is twice the group luminosity of RR~242. Since the number of {\it giant} galaxies is the same in both systems ($N_{giant}=10$), the individual giants of RR~216 contain more (luminous) mass than the ones in RR~242. The $M/L$ ratios are based on $M_{vir}$, motivated by the common use of $M_{vir}$ in the literature. The $M/L$ ratios of the groups are consistent with typical values found for poor groups of galaxies \citep[see e.g.,][]{Firth06}, although at the lower limit, indicating that the virial mass is indeed underestimating the total mass of the systems (as expected by the definition of $M_{vir}$). 

The positions of group members within our E+S systems is presented in Fig.~\ref{fig16}. The groups are moved to a common distance: the field of view shown in the figure corresponds to $\sim$ 2.5 $h_{100}^{-1}$ Mpc on each side. We plot the large-scale environment search radius of 90$^\prime$ (see Appendix~A) centred on the E member (solid line), the WFI field of view (central square), the 0.5 $h_{100}^{-1}$ Mpc radius (dashed line), and the mean harmonic radius $R_H$ (dotted line), the latter two centred on the optical group centre. The central, solid-line circle is $R_H$ calculated only for members inside 0.5 $h_{100}^{-1}$ Mpc and centred on the respective group centre. The plot makes clear that the optical group centres do not coincide with the positions of the pairs except for RR~210. This group also appears to be the most compact system with small $R_H$. This may be caused by the higher number of galaxies with intermediate luminosity identified by NED (due to the proximity of the group). This system also involves the pair with the smallest projected separation, which also causes the luminosity-weighted $R_H$ to be smaller. The unweighted $R_H$ is comparable to that of the other groups (see Table~\ref{table5} for values). The galaxies around RR~216 are spread over the full area investigated without any central concentration, which is indicative of the higher large-scale density of its environment (outskirts of the Hydra-Centaurus Cluster).

The spatial distribution of faint members can only be investigated for the group centres (within the WFI field of view $\sim$ 250 $h_{100}^{-1}$ kpc radius around the E member). The redshift information of galaxies outside this field is taken only from NED, which is highly incomplete at fainter magnitudes. The radial density profile of the faint group members ($M_R > -19 +5 \log h_{100}$) is plotted in Fig.~\ref{fig17} for all four groups. The number of group members is counted in radial bins of 20 $h_{100}^{-1}$ kpc width and divided by the area of that bin. The radius is measured from the field centre, i.e., the E member. The first bin 0 - 20 kpc is not plotted because no galaxies are found there because of the large extent of the bright elliptical. The effective radius of the smallest of the four Es is $\sim$ 12 kpc. The faint galaxies in RR~242 are clearly concentrated around the bright elliptical, while around RR~143 there is no central concentration. The situation for the other two pairs is unclear due to the high and asymmetrical incompleteness.  In Fig.~\ref{fig17}, the radial density profiles of the faint members of our E+S systems are compared with the profile measured for the isolated elliptical and fossil group candidate NGC~1132 and the composite profile of five X-ray bright groups \citep[see Fig. 3 in][]{Mul99}. In contrast to our radial density profiles that are computed using only spectroscopically confirmed member galaxies, these two last profiles were compiled by assuming that all the detected faint galaxies in each group field belong to the group. Since the background galaxies are not expected to concentrate around the bright elliptical but instead be uniformly distributed in the field, we expect a roughly constant shift in number density throughout the field. This is indeed the case for our richest system RR~242, in which the radial density profile is similar to that of the five \citet[][]{Mul99} groups.

\begin{figure*}[t]
\includegraphics[width=9.2cm]{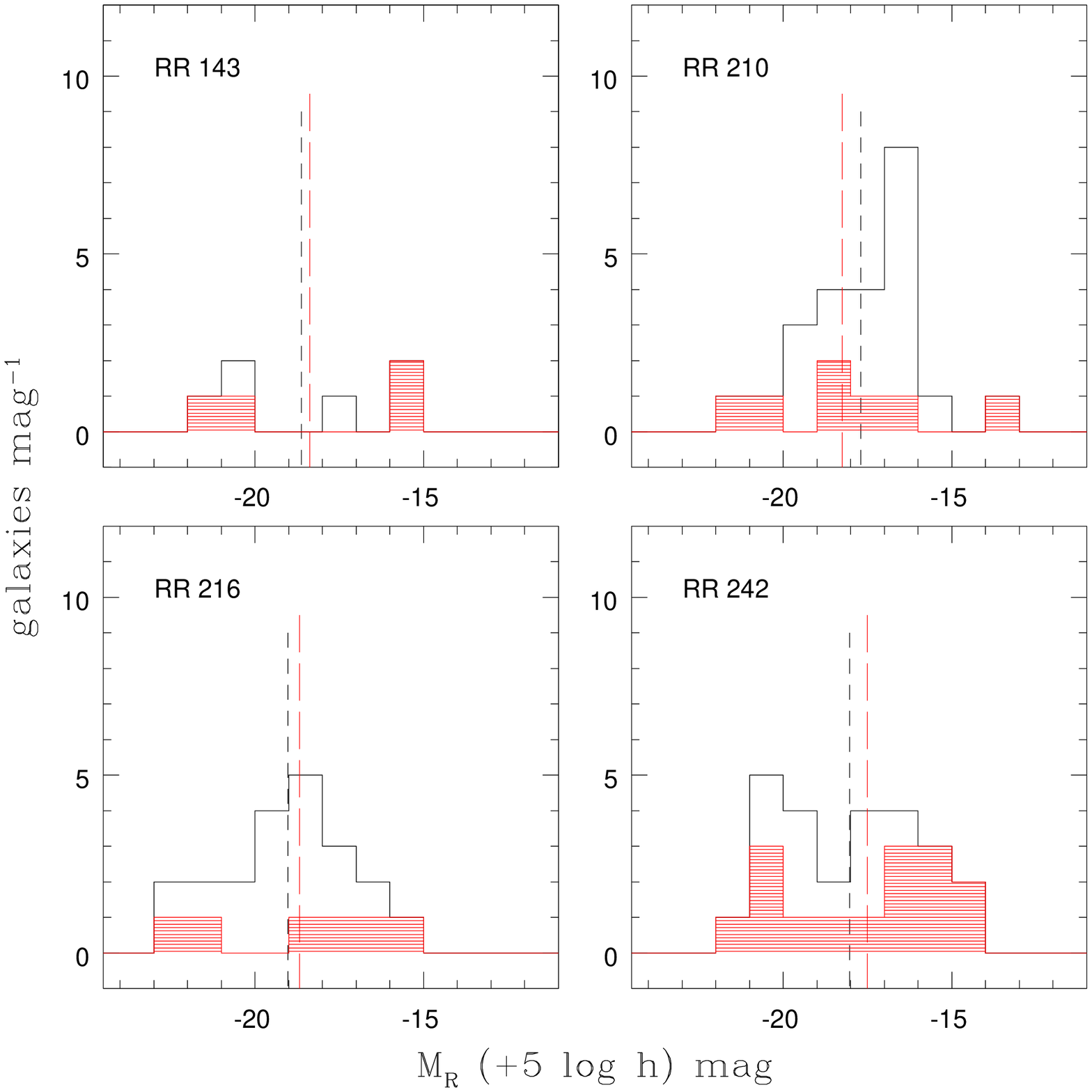}
\includegraphics[width=9.2cm]{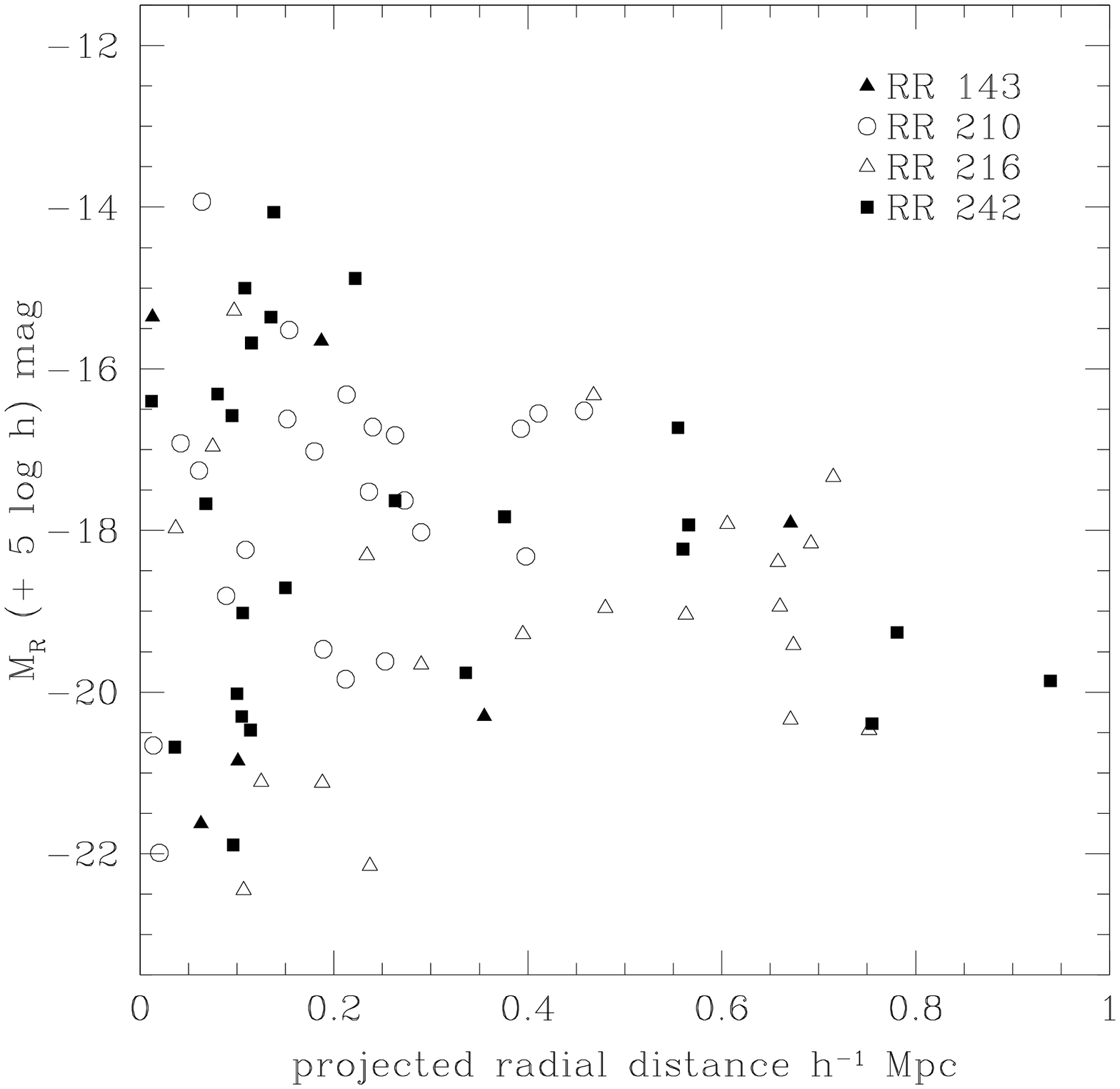}
\caption{Distribution of absolute magnitudes. Left: Absolute magnitude histogram of members of the 4 groups in the WFI field (red) and in the 90$^\prime$ radius. The vertical dashed lines indicate the mean absolute magnitude of group members within the WFI field (red, long dashed) and the 90$^\prime$ radius (black, short dashed). Right: Absolute magnitude vs. projected radial distance from the group centre. Galaxies of each group are marked with different symbols: solid triangles: RR~143; open circles: RR~210; open triangles: RR~216; solid squares: RR~242.
\label{fig18}}
\end{figure*}

\subsection{The E+S system luminosity function and dwarf-to-giant ratio}

To simplify comparison with luminosity functions in the literature, absolute magnitudes are computed in $+5 ~\log ~h_{100}$ mag. We consider separately the distributions of absolute magnitudes for all member galaxies: a) within the WFI field of view, and b) within the 90$^\prime$ radius (see Appendix~A). OLFs are computed with the completeness-corrected galaxy counts only within 0.5 $h_{100}^{-1}$ Mpc in order to compare the same physical region in all groups. Figure~\ref{fig18} shows the distribution of absolute magnitudes in the four groups, both for members located within the WFI field of view and for all members within a 90$^\prime$ radius as given in Table~\ref{tableA1}. The mean absolute magnitudes of the WFI field subsamples do not differ significantly from the entire group values.  The values are consistently around $M_R \sim$ -18 mag apart from RR~216, which has a higher value due to both its brighter elliptical pair member and other very bright members in the larger scale environment.

It is remarkable that very few faint members were found.  Confirmed companions tend to show intermediate luminosities in the supposed transition region from ``normal'' to dwarf galaxies ($M_R \sim -18$ \citep{Fergusson94}). The intermediate region tends to be populated by faint S0s, spirals, and dwarf ellipticals. Despite the high number of very faint candidates (the median of all 4 candidate samples is fainter than $M_R \sim -14 + 5 \log h_{100}$), no faint dwarf-irregular galaxies were found. 

The dependence of galaxy magnitude on position within a group is also investigated in Fig.~\ref{fig18}. The trend is that the brightest galaxies in the groups are more centrally concentrated than the galaxies of intermediate luminosity, as already discussed in the analysis of peculiar velocities (Fig~\ref{fig10}). In contrast to RR~143, RR~216 and RR~242 contain some brighter members outside $\sim$ 0.6 $h_{100}^{-1}$ Mpc. This region lies outside the investigated area for RR~210. The lack of faint galaxies outside $\sim$ 0.2 $h_{100}^{-1}$ Mpc is easily explained since it corresponds to the approximate size of the WFI field of view over which we searched for faint member galaxies.

\begin{figure*}
\includegraphics[angle=270,width=18.5cm]{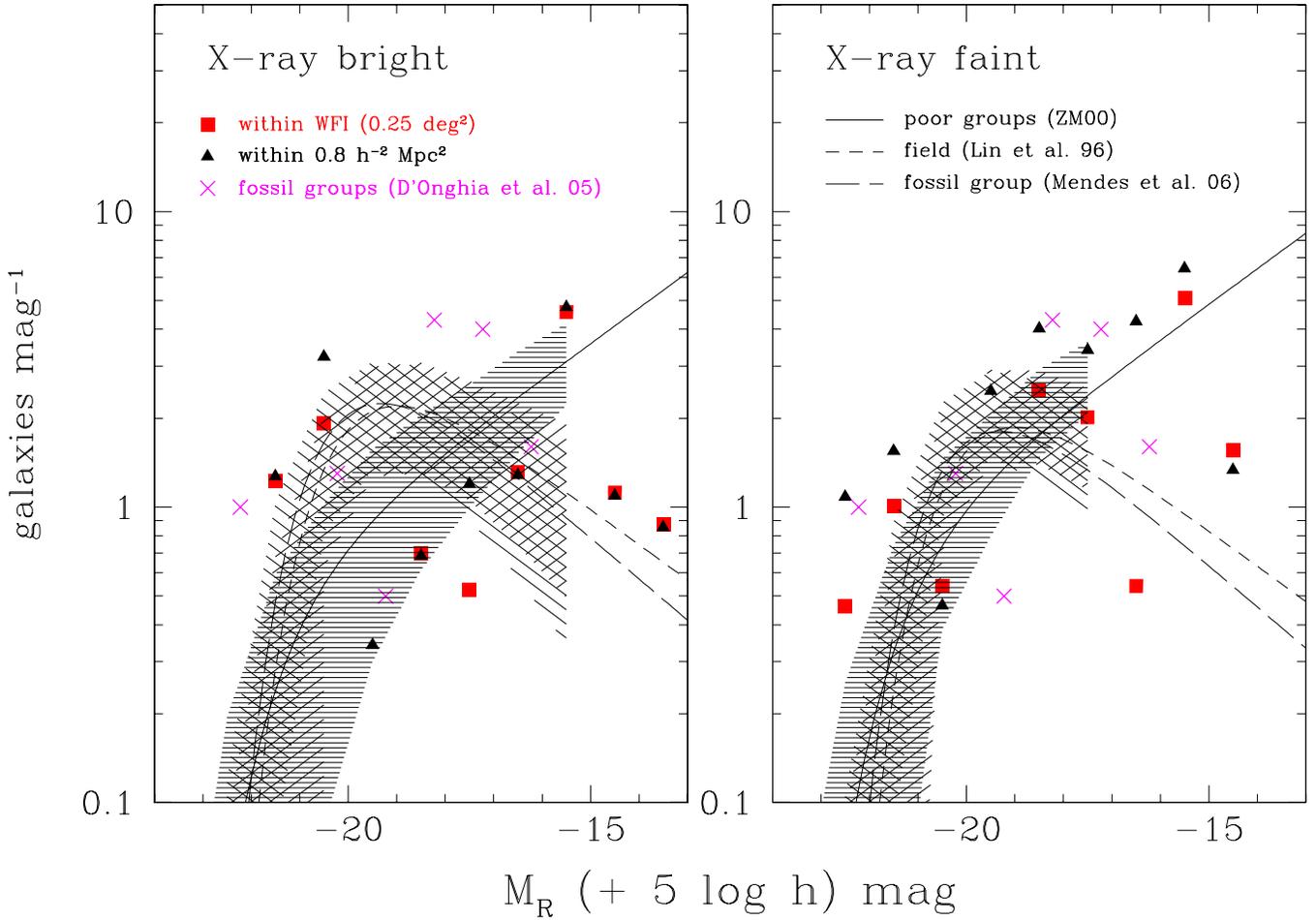}
\caption{Optical Luminosity Functions (OLF) of the 4 groups. The OLFs of the 2 X-ray bright and the 2 X-ray faint groups are combined respectively. The OLFs are computed for the large scale sample (black triangles) and the WFI field subsample (red squares). Note that outside the WFI field the radial velocity information comes from NED and is highly incomplete at fainter magnitudes. The magenta crosses show the OLF of a sample of simulated fossil groups from \citet{D'Ong05}. The solid line is the OLF found for a sample of X-ray bright poor groups by \citet{ZM00}. The short dashed line shows the OLF of the local field \citep{Lin96}, whereas the long dashed line is the OLF of a fossil group observed by \citet{Mendes06}. All OLFs use spectroscopically confirmed members only and are completeness corrected in a similar way as our counts. The shaded regions around each OLF shows the $1 \sigma$ deviations expected due to our low number statistics obtained from a set of Monte Carlo Simulations (see text).
\label{fig19}}
\end{figure*}

Figure~\ref{fig19} shows the combined OLFs of the two groups RR~143 and RR~242 with extended hot IGM (labelled X-ray bright in the figure) and the two groups RR~210 and RR~216, which are X-ray faint. The galaxy counts per magnitude are completeness-corrected but not normalised to the surveyed area. To compare the same physical area in all groups, only galaxies within 0.5 $h_{100}^{-1}$ Mpc projected distance from the pair elliptical were considered (the area sampled in the closest group RR~210). Before combining them, the OLF of each group was normalised to have the same number of members brighter than a certain magnitude. We used $M_R = -15 +5 \log h_{100}$ for the X-ray bright groups, since the spectroscopic completeness was higher than 50\% in all bins down to this magnitude. For the X-ray faint groups, only the numbers above $M_R = -17 +5 \log h_{100}$ are used due to the high incompleteness in RR~216 below $M_R = -17 +5 \log h_{100}$ ($<$ 10\% completeness, see Fig.~\ref{fig4}). The OLFs of both groups were then averaged. This procedure ensures that the shape of the combined OLF is not weighted towards the richer group. 

An analytic form of the luminosity function fitted to the observed distribution of magnitudes was described by \citet{Schechter76}. It has a steeply rising slope at the bright end that then levels off or even decreases at fainter magnitudes. It has 3 parameters: $M^\ast$, the transition between bright and faint end slope, $\alpha$, the faint end slope, and $\Phi^\ast$, the galaxy density at $M^\ast$, which gives the normalization of the OLF. This latter parameter gives the number of galaxies per Mpc$^{3}$ and is difficult to compute, since it requires a survey covering an extensive area. Since it does not affect the shape of the OLF, it is often just adjusted to fit the number of observed galaxies. Our spectroscopic completeness drops below 50\% for magnitudes fainter than $M_R = -15 +5 \log h_{100}$. Therefore, we decided to adjust the OLFs from the literature to our number of galaxies brighter than $M_R = -15 +5 \log h_{100}$. Galaxies fainter than this limit are not considered further here. For the X-ray faint groups, the OLFs are adjusted to the galaxy number brighter than $M_R = -17 +5 \log h_{100}$. Since we lack information about faint galaxies outside the WFI field of view, we compared the OLFs from the literature to our galaxy numbers within the WFI field (red squares). For comparison, the counts for all galaxies within a 0.5 $h_{100}^{-1}$ Mpc radius ($\sim$ 0.8 $h_{100}^{-2}$ Mpc$^2$) are also plotted (black triangles). The OLF of a sample of simulated fossil groups of galaxies (into which our pairs could evolve) is taken from \citet{D'Ong05} (magenta crosses). The solid line represents the OLF for X-ray detected poor groups found by \citet{ZM00}. They investigated a sample of five poor galaxy groups with X-ray halos of comparable luminosity to our two X-ray bright pairs and found that their OLF can be fitted by a Schechter function with $M^\ast = -21.6 +5 \log h_{100}$ and $\alpha = -1.3$. These values are comparable to those measured for nearby rich cluster OLFs \citep[see e.g.,][]{Trentham96,Driver98}. The short-dashed line shows the OLF of the local field from the Las Campanas Redshift Survey (LCRS hereafter) found by \citet{Lin96} with $M^\ast = -20.29 +5 \log h_{100}$ and $\alpha = -0.7$.  LCRS results agree closely with those for a large sample of the most isolated galaxies in the northern sky (AMIGA sample, Verdes-Montenegro et al. 2005). Finally, the long-dashed line represents the OLF of the spectroscopically confirmed members of a fossil galaxy group from \citet{Mendes06}.

The possible effects of incompleteness and low number statistics were simulated again using the Monte Carlo method. A set of 1000 fake groups with the same number of members as our combined group samples and whose members follow a certain Schechter OLF was constructed. To do this, a random pair of values in the plane of galaxy magnitude versus galaxy number was produced. If this data point happened to lie under the curve of the respective OLF, this galaxy magnitude was included in the group sample. This procedure continued until the desired number of members was reached and then repeated for the next group. Finally, the magnitude histogram was computed for each group and the mean and rms of all group histograms were computed in each bin. This was done for the group, field and fossil group OLFs with the number of members of both, the X-ray-bright (left panel) and the X-ray faint (right panel) combined group. Only galaxies down to $M_R = -15 +5 \log h_{100}$ in the X-ray-bright and $M_R = -17 +5 \log h_{100}$ in the X-ray faint groups were considered. The expected $1 \sigma$ deviation of each OLF is shaded, horizontally for the group OLF \citep[][ZM00 hereafter]{ZM00}, and diagonally for the field \citep{Lin96} and fossil group (FG) \citep{Mendes06} OLFs (which are quite similar).

The combined OLF of the two X-ray bright groups shows evidence of a non-Schechter form with an excess of bright galaxies or a pronounced dip at $M_R \sim -20 +5 \log h_{100}$. The same dip is present in the X-ray faint group OLF (red squares), although it disappears at larger radii, outside the WFI field of view (black triangles). To test the significance of this dip in the X-ray bright groups with respect to the X-ray faint groups, we compared the counting errors of the X-ray bright and X-ray faint groups in each magnitude bin (black triangles). The difference in counts between the two OLFs exceeds the sum of their errors in the bins from $M_R =$ -20.5 to -18.5 $+5 \log h_{100}$, which suggests that they indeed differ. For the sake of clarity, we do {\it not} show the counting errors for each bin in Fig.~\ref{fig19}.

The X-ray bright OLF is inconsistent with the OLF of X-ray detected poor groups found by ZM00 (at more than 1$\sigma$). The LCRS field OLF is also unable to fit our observations because there are not enough giants of intermediate luminosity (below $M_R \sim -20 +5 \log h_{100}$). The X-ray faint group OLF on the other hand is consistent with the OLF of the ZM00 groups, although there remains an excess of bright giant galaxies. However, within the rms expected from the Monte Carlo simulations, the X-ray faint group OLF is consistent with the OLF of the ZM00 groups.

Interestingly, the OLF of the simulated and observed fossil groups (FGs) are quite different, although this difference is based on only one observed group. The OLF of the sample of simulated FGs shows a prominent gap between -20 $\leq M_R \leq $-19 + 5 $\log h_{100}$ mag, whereas the observed FG reaches its maximum in this range. A similar gap in the OLF of X-ray faint groups was also found by \citet{Miles04}, however, their group membership is based on galaxy colours and not on spectroscopy as in this study. The same lack of galaxies is present in our groups, both X-ray bright and faint. In the X-ray faint groups however, this gap vanishes if a larger area is considered (black triangles in Fig.~\ref{fig19}), similar to the findings of \citet{Miles06}. They found that the gap in the {\it infrared} LF of X-ray faint groups is vanishing at larger radii. However, the gap in our combined X-ray bright group OLF remains and is not an effect of the small survey area. The OLF of the observed FG seems to drop at fainter magnitudes, although the authors indicate that this may be due to their limiting magnitude of $M_R \sim$ -18 + 5 $\log h_{100}$ mag and may actually be rather a dip at this magnitude than a continuing drop towards fainter magnitude. This makes the FG OLF consistent with our X-ray faint group OLF, although it cannot be distinguished from the other two OLFs with our low number statistics, as shown by the results of the Monte Carlo simulations (shaded areas).

Incompleteness effects are certainly an issue in our sample, although they are not expected to play a major role down to $M_R \sim -17 +5 \log h_{100}$. Apart from 3 galaxies in the range $-19 \leq M_R \leq -17$ that were missed in RR~143, all candidates without measured redshift above $M_R = -17$ are accounted for by the completeness correction. If these 3 missed candidates were members of RR~143, the OLF of the combined X-ray bright groups would be slightly higher in this range and closer to the ZM00 OLF. However, this would  influence neither that gap above -19, nor the excess of bright galaxies relative to the other OLFs, and it would not reach into the $1 \sigma$ area of the field and fossil group OLFs. So our combined X-ray bright group OLF would still significantly differ from the OLFs in the literature.

\begin{table}
\begin{tiny}
\caption{Dwarf-to-giant (D/G) ratios obtained from the OLFs.}
\begin{tabular}{lcccc}
\hline\hline
group & D/G$_{0.5}$ & D/G$_{0.5,17}$ & D/G$_{WFI}$ & D/G$_{WFI,17}$ \\
\hline
RR~143 & 1.3$\pm$0.7 & 0.0$\pm$0 & 1.9$\pm$1.2 & 0.0$\pm$0 \\
RR~210 & 4.2$\pm$1.4 & 1.8$\pm$0.7 & 3.9$\pm$2.1 & 2.0$\pm$1.2 \\
RR~216 & 3.0$\pm$1.0 & 1.0$\pm$0.4 & 8.0$\pm$3.8 & 2.5$\pm$1.4 \\
RR~242 & 3.4$\pm$1.1 & 0.9$\pm$0.4 & 4.3$\pm$1.6 & 0.9$\pm$0.5 \\
Combined X-ray bright & 2.2$\pm$0.6 & 0.4$\pm$0.2 & 3.0$\pm$0.9 & 0.4$\pm$0.2 \\
Combined X-ray faint & 3.5$\pm$0.8  & 1.3$\pm$0.4 & 5.8$\pm$2.0 & 2.2$\pm$0.9 \\
\hline
 & & & & \\
\end{tabular}
\label{table6}
\footnotetext{}{
Column 2: 
D/G$_{0.5}$: dwarf-to-giant ratio for galaxies within 0.5 $h_{100}^{-1}$ Mpc.\\
Column 3: D/G$_{0.5,17}$: dwarf-to-giant ratio for galaxies within 0.5 $h_{100}^{-1}$ Mpc using only
dwarfs brighter than $M_R = -17 +5 \log h_{100}$.\\
Column 4: D/G$_{WFI}$: dwarf-to-giant ratio for galaxies within the WFI field of view.\\
Column 5: D/G$_{WFI,17}$: dwarf-to-giant ratio for galaxies within the WFI field of view
using only dwarfs brighter than $M_R = -17 +5 \log h_{100}$.}
\end{tiny}
\end{table}

Dwarf-to-giant (D/G) ratios are determined from the completeness-corrected counts for both individual groups and for the two combined samples. The limit
between giants and dwarfs is again $M_R = -19 +5 \log h_{100}$ following \citet{ZM00}. The values are computed for the full dwarf sample and for galaxies brighter than $M_R = -17 +5 \log h_{100}$, so as to compare with literature values. We also computed D/G for all galaxies within the WFI field of view since outside this field information about faint member galaxies is very poor. The errors are the combined counting errors of each dwarf and giant sample. All values are given in Table~\ref{table6}. An average D/G value for the \citet{ZM00} X-ray luminous poor groups (down to $M_R = -17 +5 \log h_{100}$) is 1.9 $\pm$ 0.4 compared to a value of less than 1.0 for local group galaxies in this magnitude range \citep{Pritchet99}. Our X-ray faint groups have higher D/G ratios than our X-ray bright groups. The value of the combined X-ray faint E+S systems is close to the value found for the \citet{ZM00} X-ray luminous poor groups  (D/G$_WFI,17 = 2.2\pm0.9$), whereas the D/G of our X-ray bright systems is significantly lower (D/G$_WFI,17 = 0.4\pm0.2$). By considering the two X-ray bright E+S systems to be the groups with more complete coverage, the result is unlikely to be caused mainly by incompleteness effects. However, in the field of RR~143, three candidates in the range  -19 $+5 \log h_{100}<M_R<-17 +5 \log h_{100}$ (see Fig.~\ref{fig3}), were not observed with VIMOS. Assuming that these 3 galaxies were all members of the E+S system, the D/G would reach a value of $\sim$ 1, similar to that of RR~242. This value is still lower than the D/G of 1.9 from \citet{ZM00}, as well as the D/G ratio found for our X-ray faint E+S systems.

\subsection{What maps the hot inter-galactic medium?}

Groups with an extended hot IGM usually have a giant elliptical that is typically the brightest group member and located near or at the peak of the ``smooth, symmetric'' X-ray emission \citep[see e.g., the review in][]{M00}. Another distinctive characteristic of groups with a hot IGM component was found by \citet{ZM98}. By means of multi-object spectroscopy, they found a significantly higher number of faint galaxies ($\sim$ 20 - 50 members down to magnitudes as faint as $M_B \sim -14 + 5 \log h_{100}$ to -16 $+ 5 \log h_{100}$) in groups with a significant amount of hot IGM. 

Analysis of the faint population in our E+S systems suggests that the presence of extended diffuse X-ray emission is not necessarily connected to the presence of such a numerous faint galaxy population. At the same time, our E+S systems appear to be gravitationally bound structures. The velocity dispersion profile suggests that the dispersion is not constant with radius. It shows a maximum at $\sim$ 0.2 $h_{100}^{-1}$ Mpc and decreases until $\sim$ 0.5 $h_{100}^{-1}$ Mpc before increasing again at higher projected radii from the optical group centre. This suggests that the groups have a dynamical boundary at $\sim$ 0.5 $h_{100}^{-1}$ Mpc. Furthermore, dynamical analysis of the four E+S systems indicates that the pair is displaced from the optical group centre, suggesting that the hot IGM is a ``local'' phenomenon. The X-ray emission is centred on the E member of the pair indicating that it may be the principal reason for the presence of the hot X-ray emitting gas.

In Paper~III, we showed that RR~143 and RR~242 have a luminous, extended, hot IGM, while both RR~210 and RR~216, although of similar ``optical'' and kinematical characteristics, are X-ray underluminous with respect to other loose groups or mature Es, if we consider the emission connected with the elliptical galaxy. The diffuse X--ray emission from the hot intra-group medium (IGM) detected in compact but also in loose, poor groups, has often been taken as a direct measure of the group potential \citep[see e.g.,][]{M00}. At the same time, \citet{Sansom00} offered another interpretation of the large spread in X-ray luminosity among ETGs. They suggest that ellipticals, and in general early-type galaxies, showing fine structure -- such as e.g., shells and dust-lanes -- tend to have a fainter X-ray luminosity, although the dispersion is very large. These early-type galaxies are considered {\it dynamically young} i.e., showing evidence of recent accretion/merging events.

Several young systems, although in small groups, do not show extended emission and lack a substantial group component \citep{Osul01}. XMM-$Newton$ and $Chandra$ observations found three underluminous elliptical galaxies (NGC~3585, NGC~4494, and NGC~5322) all of which show evidence of recent dynamical disturbances, including kinematically distinct cores as in the case of NGC~474 \citep{Hau96}. \citet{Rampazzo06} found that the X-ray luminosity of NGC~474, the early-type member of another E+S system (Arp 227) in the northern hemisphere, lies about two orders of magnitude below that of dominant group members and is located in the area of the log~$\rm L_B$ - log~$\rm L_X$ plane where the X-ray emission could be explained by the superposition of discrete X-ray sources. 

The position of interacting or post-interacting galaxies, such as those exhibiting fine structures or kinematical perturbations, in the L$\rm_X$ - L$\rm_B$ plane is consistent with the hypothesis that their X-ray emission comes from discrete sources only, although their L$\rm_X$/L$\rm_B$ ratios are not as low as that of NGC~474. \citet{Sansom00} and \citet{Osul01} interpreted the negative trend observed between L$\rm_X$/L$\rm_B$ (linked to the gas content in early-type galaxies ) and the morphological disturbance quantified by the fine-structure parameter $\Sigma$ (linked to the age/dynamical stage) as evidence that several gigayears are required to accumulate hot, gaseous halos, so that recent mergers/young systems are deficient in hot gas. \citet{Osul01} also attributed some of the scatter seen in the global L$\rm_X$ versus L$\rm_B$ relation to the evolutionary stage and past merger history of early-type galaxies. \citet{Brassington07} studied the X-ray emission of nine merging systems believed to represent different phases of the merging process. They suggested that (1) the X-ray luminosity peaks $\sim$300 Myr before nuclear coalescence; (2) at a time $\sim$1 Gyr after coalescence, the merger remnants are fainter compared to mature ellipticals; while (3) at a greater dynamical age ($\geq$3 Gyr) remnants start to resemble typical ellipticals in their hot gas content. On these grounds, the above authors support the idea that a {\it halo regeneration} takes place within low L$_X$ merger remnants.  

A possible explanation of the diverse X-ray properties of our E+S systems could then be connected with the dynamical age of the dominant E galaxy. Groups with similar environments will be in different evolutionary phases that can be traced by the giant central elliptical galaxy: interaction (accretion/merging) disrupts the hot gas halo, which is then built up again during ongoing evolution on a timescale of a few gigayears. This implies that the majority of the X-ray emission is unlikely to be associated with the group potential, as also suggested by the kinematics and dynamics of these systems. In this scenario, RR~242 would be the most evolved system, and RR~216, which is expected to have a significant number of faint companions and faint X-ray emission might be experiencing an active phase of dynamical evolution. The same might be valid for RR~210, which, despite its apparent lack of faint companions, is embedded in a compact larger-scale structure (in projection and in redshift space). The apparent lack of faint companions concentrated around the E+S pair might be due to incompleteness. Another possible explanation of the relative X-ray faintness of these two systems is that they are still in the process of collapse, so the IGM has yet to be compressed and heated to X-ray temperatures \citep{Ras06}. However, this seems unlikely regarding the short crossing times of the pairs, arguing for a virialization at least of the group centres. The clear signatures of interaction found in the two pairs, including the presence of diffuse light suggests a long-lasting coevolution of the pair galaxies. The case of RR~143 is quite different with a real lack of faint galaxies. It is interesting that the scarcity of faint members continues on larger scales. We find only two other galaxies of similar redshift within $\sim$ 1 $h_{100}^{-1}$ Mpc.

\section{Summary and conclusions}

We have presented VLT-VIMOS observations in a search for faint galaxy members of four E+S systems. Candidate members were identified by applying photometric criteria to WFI images covering a field of view of about 0.2 $h_{100}^{-1}$ Mpc radius (see section~\ref{obsandred}). We investigated the morphological and photometric characteristics of the new group members as well as their spectral properties. We used the new data to determine the group dynamics as well as the combined group luminosity functions.

We found the following results:

\newcounter{count1}
\begin{list}{\arabic{count1}. }{\usecounter{count1}}

\item Two and ten new members are confirmed for RR~143 and RR~242, respectively. We found two and three new members associated with RR~210 and RR~216, respectively, which are both only partially covered by our VIMOS observations. The new members increase the galaxy populations to 4, 7, 6, and 16 members in a 0.5$^\circ$ $\times$ 0.5$^\circ$ field around the pairs RR~143, RR~210, RR~216, and RR~242, respectively, down to $M_R \sim$ -12 $+ 5 \log h_{100}$. We applied incompleteness corrections that were necessary for the subsequent investigations.\\

\item The morphological study of the new members based on our detailed surface photometry indicated a high fraction of S0 galaxies (40\%), most with low B/T ratios. A morphology-radius relation is apparent for the combined group sample. Galaxies with higher B/T ratios appear to be more concentrated towards the field centres (i.e., towards the E member of the pair), although there is a large spread of B/T ratios. However, the relation between morphology and local projected number density seems to be more significant (at the 95\% confidence level) arguing for the presence of a morphology-density relation. This suggests that the very local environment has a strong influence on galaxy morphologies and is responsible for shaping our faint galaxies. Signatures of interaction and merging are found in the group sample. Asymmetries, filaments, and shells are detected in several galaxies.\\

\item Spectra of the new members are indicative of an old stellar population in the vast majority of galaxies. No blue objects, dIrr or tidal dwarfs are present in our sample. This is also supported by the colour-magnitude relation for group galaxies. Although we reach into the domain of dwarf irregular (dIrr) and spheroidal (dSph) galaxies that we find in the Local Group, many of them would likely be below our detection limit.\\

\item Our dynamical analysis indicates short crossing times for all systems suggesting that at least the centres of the groups are virialised. The E pair members dominate the groups: in RR~143 and RR~210, they represent $\sim$ 1/2 and in RR~216 and RR~242 $\sim$ 1/3 of the total group light. The dynamical quantities appear uncorrelated with group X-ray luminosity. RR~242 has the highest velocity dispersion and virial mass, while RR~143 has the lowest (both are X-ray bright). The harmonic and virial radius of the groups are similar for RR~143, RR~210, and RR~242. In RR~216, member galaxies are less concentrated towards the group centre. This may be due to the different large-scale environment of this pair, which is in the outskirts of the Hydra-Centaurus cluster region. 

The velocity dispersion seems to vary with distance from the optical group centre. The pair ellipticals being the brightest group galaxies are more centrally concentrated than giants and dwarfs, but still about 0.1 $h_{100}^{-1}$ Mpc from the group centre (apart from RR~210, which is located precisely at the optical group centre). This also means that the hot IGM, which is centred on the elliptical, is shifted from the optical group centre. The X-ray emission then seems to be connected with the bright elliptical of the pair and its evolutionary phase rather than with the group environment.\\

\item The OLF of X-ray bright E+S systems differs from the OLF of X-ray faint systems, in line with the findings of previous authors \citep{Miles04}.
The X-ray bright E+S system OLF also differs from both that of the sample of X-ray luminous poor groups in \citet{ZM00}, and the OLF of the local field and isolated galaxies \citep{Lin96}. This comparison suggests that the OLF of poor X-ray detected galaxy systems is not universal, in contrast to the results of \citet{ZM00}. Despite the environmental differences between the two X-ray luminous groups in our sample, their normalised OLFs are quite similar, showing a lower D/G ratio (compared to the groups in \citet{ZM00}) with equal numbers of dwarfs and giants. Within the giant regime ($M_R$ brighter than $-19 +5 \log h_{100}$), our X-ray bright groups also show an interesting behaviour: they have a higher number of bright galaxies or, to put it in a different way, they lack galaxies between -20 $\leq M_R \leq $ -19 + 5 $\log h_{100}$ mag. Their OLF is reminiscent of the luminosity function of NGC~5846, which is a group considered to be dynamically evolved \citep{Mahdavi05}. The X-ray bright OLF is also comparable to the OLF of a sample of simulated fossil groups \citep{D'Ong05} showing a similar gap around $M_R \sim$ -19 + 5 $\log h_{100}$ mag. This could indicate that our X-ray bright E+S systems are more dynamically evolved than the \citet{ZM00} groups and that the E pair members are the remnants of this evolution. 
 
On the other hand, the OLF of the X-ray faint E+S systems agrees with the OLF of the \citet{ZM00} groups and these systems, furthermore, have a similar D/G ratio ($\sim$ 2). The X-ray faint group OLF also agrees very well with the spectroscopically confirmed OLF of an observed fossil group \citep{Mendes06}. The X-ray faint groups may thus be a phase in the dynamical evolution of the \citet{ZM00} groups where the recent or ongoing interaction, in which the E member is involved, could have destroyed or at least decreased the luminosity of the IGM. The X-ray halo could then be built up again during the subsequent passive evolution of the elliptical \citep[e.g.,][]{Sansom00,Brassington07}. Their OLF would be consistent with these systems evolving into fossil groups similar to that observed by \citet{Mendes06}. \\

\end{list}

Interaction-induced rejuvenation episodes may be present in a small fraction of our sample as suggested by the presence of fine-structures. The E members in both RR~210 and RR~216 are also good candidates for showing rejuvenation signatures in their stellar population. This finding would reinforce our hypothesis that their faint X-ray emission is connected with the phase of their dynamical evolution.  The presence of a young stellar population in both the giant and faint galaxy members can be ascertained by the study of absorption line-strength indices that will be carried out in a future paper.

\begin{acknowledgements}
We thank the anonymous referee for many useful suggestions that certainly improved the content of this paper. RG wants to thank Jes\'us Varela for many helpful discussions and his assistance in preparing the OLFs. RG and RR thank the INAF-Osservatorio Astronomico di Padova and the Institut f\"ur Astronomie Universit\"at Wien, respectively, for the kind hospitality during the paper preparation. RG, RR and WWZ acknowledge the support of the Austrian and Italian Foreign Offices in the framework science and technology bilateral collaboration (project number 25/2004). RG and WWZ acknowledge the support of the Austrian Science Fund (project P14783). RR and GT acknowledge the partial support of the Agenzia Spaziale Italiana under contract ASI-INAF I/023/05/0. This research has made use of the NASA/IPAC Extragalactic Database (NED) which is operated by the Jet Propulsion Laboratory, California Institute of Technology, under contract with the National Aeronautics and Space Administration. The Digitized Sky Survey (DSS) was produced at the Space Telescope Science Institute under U.S. Government grant NAG W-2166. The images of these surveys are based on photographic data obtained using the Oschin Schmidt Telescope on Palomar Mountain and the UK Schmidt Telescope. The plates were processed into the present compressed digital form with the permission of these institutions. This research has made use of SAOImage DS9, developed by Smithsonian Astrophysical Observatory.
\end{acknowledgements}

\begin{appendix}
\section{Properties of group members used for dynamical calculations}

Since our WFI observations probably do not cover the complete extension of each group, we searched for additional group member galaxies in the environment of each pair outside the WFI field (see also Paper~III). The NASA/IPAC Extragalactic Database (NED) was used for this search. The area investigated was 90\arcmin~  corresponding to $\sim$ 1 $h_{100}^{-1}$ Mpc for the farthest pair. We also chose this radius to mimic the observed area of \citet{ZM98}, the group sample to which we compared our results. The maximum velocity difference within which a galaxy was considered to be a group member was chosen to be $\Delta v \leq 1000$ km s$^{-1}$.

Previous authors have proposed a luminosity-weighted formulation to calculate the group dynamics \citep[e.g.,][]{Fergusson90,Firth06}. To investigate the differences introduced by weighting, we calculated the group dynamics by using both luminosity weighting and uniform weights ($w_i = 1$).

In the luminosity-weighted approach, each galaxy was weighted by its relative luminosity calculated from the galaxy's R-band magnitude
\begin{equation}
w_i = 10^{-0.4~m_{Ri}}
\end{equation}
The coordinates of the optical group centre $\alpha_{group}$ and $\delta_{group}$ are given by averaging the luminosity-weighted coordinates $\alpha_i$ and $\delta_i$ of all group members
\begin{equation}
\alpha_{group} = \frac{\sum_{i}  \alpha_i~w_i}{\sum_{i} w_i}   ~~and~~ \delta_{group} = \frac{\sum_{i}  \delta_i~w_i}{\sum_{i} w_i}
\end{equation}
The luminosity-weighted mean velocity, i.e., the velocity of the optical group centre was calculated similarly
\begin{equation}
v_{group} = \frac{\sum_{i}  v_i~w_i}{\sum_{i} w_i}
\end{equation}

The luminosity-weighted line-of-sight velocity dispersion was calculated by summing over the weighted squared-deviation of each group member from the group velocity
\begin{equation}
\sigma_r = \left[\frac{\sum_{i} w_i ~(v_i - v_{group})^2}{{\sum_{i} w_i}}\right]^\frac{1}{2} \\
\end{equation}

The mean harmonic radius is a measure of the compactness of the group and was calculated from the projected separations $R_{ij}$ between the i-th and j-th group member
\begin{equation}
R_H =  \left[\frac{\sum_{i} \sum_{j<i} (w_i w_j)/ R_{ij}}{\sum_{i} \sum_{j<i} w_i w_j}\right]^{-1}
\end{equation}

The virial radius was connected to $R_H$ with
\begin{equation}
R_{vir} = \frac{\pi ~R_H}{2}
\end{equation}

The crossing time $t_c$ was computed from the mean distance of group members from the optical group centre, $\langle r \rangle$, and the mean velocity relative to the group centre, $\langle v \rangle$, following the definition of \citet{Rood78}

\begin{equation}
t_c = \frac{\langle r \rangle}{\langle v \rangle}
\end{equation}

Multiplying $t_c$ with H$_0$ gives the crossing time in units of the age of the universe, independent of the choice of H$_0$.\\

Different mass estimators can be found in the literature. Following the discussion in \citet{Heisler85}, we computed the virial mass $M_{vir}$ and the projected mass $M_P$

\begin{equation}
M_{vir} = \frac{3}{G} \sigma_r^2 R_{vir}
\end{equation}

\begin{equation}
M_P = \frac{32}{\pi G} \frac{\sum_{i} w_i (v_i - v_{group})^2 r_{\perp i}}{\sum_{i} w_i}
\end{equation}

The mass-to-light ratio $M_\odot/L_\odot$ was computed from the virial mass and the group luminosity obtained by summing up the luminosities of the individual galaxies. These were calculated from the absolute magnitude of the galaxies obtained by the distance modulus given in Table~\ref{table5} and the absolute magnitude of the Sun $M_{R\odot} = 4.42$ mag \citep[taken from][]{Binney98}
\begin{equation}
L_i = 10^{0.4 (M_{R\odot}-M_{Ri})} 
\end{equation}

Table~\ref{tableA1} provides the properties of the new group members within the WFI field (ID starting with ``RR'') as well as galaxies found in the NED. All galaxies as well as only galaxies within 0.5 $h_{100}^{-1}$ Mpc from the optical group centre ($r_\perp$, Cols. 3 and 8) were used to calculate the group dynamics given in Table~\ref{table5}. The peculiar velocity $v_{pec}$ ($v_{pec} = v_i - v_{group}$) and the projected radius from the optical group centre $r_{\perp}$ were used in the analysis of the dependence of velocity dispersion on radius (Fig.~\ref{fig15}). 

The last column in Table~\ref{tableA1} gives the weights normalised to the sum over all weights, i.e., their contribution to the total luminous mass. The absolute magnitudes $M_R$ were used to calculate the luminosity $L_R$ (relative to the Sun), which was then converted into the normalised weights $wn_i$ by normalising each $L_R$ by the sum of all luminosities (i.e., the group luminosity). 

Columns 2-6 give the luminosity-weighted values, whereas uniform weights were used for values in cols. 7-11.

\begin{table*}
\begin{scriptsize}
\caption{Properties of group members used for dynamical calculations.
\label{tableA1}}
\begin{tabular}{lcccccccccc}
\hline\hline
 galaxy & ($v_{pec}$) & $r_{\perp}$ & $M_R$ & $L_R$ & $wn_i$ & ($v_{pec}$) & $r_{\perp}$ & $M_R$ & $L_R$ & $wn_i$\\
      & [km s$^{-1}$] & [Mpc] & [mag] & [$L_\odot$] & & [km s$^{-1}$] & [Mpc] & [mag] & [$L_\odot$] &\\

& \multicolumn{5}{c}{luminosity-weighted} & \multicolumn{5}{c}{uniform weights} \\
 
\hline
RR143\_09192 & -25. & 0.013 & -16.23 & 1.81E+08 & 0.0017 &  -130. & 0.095 & -16.29 & 1.93E+08 & 0.1667 \\
RR143\_24246 & 403. & 0.187 & -16.53 & 2.39E+08 & 0.0022 &   298. & 0.283 & -16.59 & 2.54E+08 & 0.1667 \\
NGC~2305 (RR~143a) & -80. & 0.063 & -22.50 & 5.84E+10 & 0.5491 &  -185. & 0.141 & -22.56 & 6.21E+10 & 0.1667 \\
NGC~2307 (RR~143b) & 194. & 0.101 & -21.72 & 2.85E+10 & 0.2677 &    89. & 0.180 & -21.78 & 3.03E+10 & 0.1667 \\
NGC 2297        & -79. & 0.355 & -21.17 & 1.72E+10 & 0.1613 &  -184. & 0.326 & -21.23 & 1.83E+10 & 0.1667 \\
ESO 087- G 050  & 217. & 0.671 & -18.78 & 1.90E+09 & 0.0179 &   112. & 0.604 & -18.84 & 2.02E+09 & 0.1667 \\
 & & & & & & & & & &\\
RR210\_11372 & -91. & 0.064 & -14.80 & 4.89E+07 & 0.0003 & -186. & 0.080 & -14.90 & 5.37E+07 & 0.0435 \\
RR210\_13493 & -10. & 0.061 & -18.13 & 1.05E+09 & 0.0067 & -105. & 0.050 & -18.23 & 1.15E+09 & 0.0435 \\
NGC~4105 (RR~210a)       & -82. & 0.020 & -22.86 & 8.19E+10 & 0.5242 & -177. & 0.007 & -22.96 & 8.99E+10 & 0.0435 \\
NGC~4106 (RR~210b)       & 132. & 0.014 & -21.53 & 2.41E+10 & 0.1540 & 37. & 0.010 & -21.63 & 2.64E+10 & 0.0435 \\
2MASX~J12063106-2951336  & 180. & 0.042 & -17.79 & 7.68E+08 & 0.0049 & 85. & 0.046 & -17.89 & 8.43E+08 & 0.0435 \\
IC~2996  & 238. & 0.109 & -19.11 & 2.59E+09 & 0.0166 & 143. & 0.110 & -19.21 & 2.84E+09 & 0.0435 \\
IC~3005  & -294. & 0.089 & -19.68 & 4.38E+09 & 0.0280 & -389. & 0.111 & -19.78 & 4.80E+09 & 0.0435 \\
2MASX J12052132-3002465 & -2. & 0.152 & -17.49 & 5.83E+08 & 0.0037 & -97. & 0.154 & -17.59 & 6.39E+08 & 0.0435 \\
6dF J1207305-301156     & 516. & 0.154 & -16.39 & 2.12E+08 & 0.0014 & 421. & 0.178 & -16.49 & 2.32E+08 & 0.0435 \\
IC 0760                 & 208. & 0.189 & -20.34 & 8.05E+09 & 0.0515 & 113. & 0.176 & -20.44 & 8.82E+09 & 0.0435 \\
2MASX J12083457-3008549 & 139. & 0.180 & -17.89 & 8.42E+08 & 0.0054 & 44. & 0.209 & -17.99 & 9.24E+08 & 0.0435 \\
ESO 441- G 004          & 300. & 0.213 & -17.19 & 4.42E+08 & 0.0028 & 205. & 0.212 & -17.29 & 4.85E+08 & 0.0435 \\
MCG -05-29-004          & 120. & 0.236 & -18.39 & 1.34E+09 & 0.0085 & 25. & 0.224 & -18.49 & 1.46E+09 & 0.0435 \\
IC 3010                 & -36. & 0.212 & -20.71 & 1.13E+10 & 0.0724 & -131. & 0.239 & -20.81 & 1.24E+10 & 0.0435 \\
2MASX J12041324-2918486 & -15. & 0.263 & -17.69 & 7.01E+08 & 0.0045 & -110. & 0.250 & -17.79 & 7.68E+08 & 0.0435 \\
6dF J1206496-302745     & 277. & 0.240 & -17.59 & 6.39E+08 & 0.0041 & 182. & 0.262 & -17.69 & 7.01E+08 & 0.0435 \\
IC 0764                 & 114. & 0.253 & -20.49 & 9.24E+09 & 0.0591 & 19. & 0.278 & -20.59 & 1.01E+10 & 0.0435 \\
ESO 441- G 014          & 128. & 0.273 & -18.50 & 1.48E+09 & 0.0095 & 33. & 0.304 & -18.60 & 1.62E+09 & 0.0435 \\
AM 1207-294 NED02       & 146. & 0.290 & -18.89 & 2.12E+09 & 0.0135 & 51. & 0.321 & -18.99 & 2.32E+09 & 0.0435 \\
ESO 440- G 044          & 180. & 0.398 & -19.19 & 2.79E+09 & 0.0178 & 85. & 0.390 & -19.29 & 3.06E+09 & 0.0435 \\
ESO 440- G 039          & 27. & 0.411 & -17.42 & 5.46E+08 & 0.0035 & -68. & 0.416 & -17.52 & 5.99E+08 & 0.0435 \\
ESO 441- G 011          & 118. & 0.393 & -17.61 & 6.51E+08 & 0.0042 & 23. & 0.403 & -17.71 & 7.14E+08 & 0.0435 \\
6dF J1203467-284015     & -111. & 0.458 & -17.39 & 5.32E+08 & 0.0034 & -206. & 0.453 & -17.49 & 5.83E+08 & 0.0435 \\
 & & & & & & & & & &\\     
RR216\_03519 & 493. & 0.097 & -16.15 & 1.69E+08 & 0.0004 & 461. & 0.124 & -16.17 & 1.73E+08 & 0.0435 \\
RR216\_04052 & -30. & 0.037 & -18.84 & 2.02E+09 & 0.0052 & -62. & 0.053 & -18.86 & 2.06E+09 & 0.0435 \\
RR216\_12209 & -526. & 0.075 & -17.83 & 7.95E+08 & 0.0021 & -558. & 0.083 & -17.85 & 8.11E+08 & 0.0435 \\
NGC~4373 (RR~216b)  & 173. & 0.107 & -23.32 & 1.25E+11 & 0.3247 & 141. & 0.118 & -23.34 & 1.27E+11 & 0.0435 \\
IC~3290 (RR~216a)   & 119. & 0.125 & -21.98 & 3.63E+10 & 0.0945 & 87. & 0.136 & -22.00 & 3.71E+10 & 0.0435 \\
ESO~322-IG~002      &  41. & 0.234 & -19.18 & 2.76E+09 & 0.0072 & 9. & 0.258 & -19.20 & 2.81E+09 & 0.0435 \\
NGC 4373A                & -288. & 0.188 & -21.99 & 3.67E+10 & 0.0954 & -320. & 0.211 & -22.01 & 3.74E+10 & 0.0435 \\
IC 3370                  & -293. & 0.237 & -23.02 & 9.47E+10 & 0.2463 & -325. & 0.238 & -23.04 & 9.66E+10 & 0.0435 \\
ESO 322- G 009           & 353. & 0.290 & -20.53 & 9.56E+09 & 0.0249 & 321. & 0.309 & -20.55 & 9.75E+09 & 0.0435 \\
ESO 321-IG 028           & -151. & 0.395 & -20.15 & 6.73E+09 & 0.0175 & -183. & 0.425 & -20.17 & 6.87E+09 & 0.0435 \\
ESO 322- G 011           &  19. & 0.480 & -19.83 & 5.02E+09 & 0.0130 & -13. & 0.464 & -19.85 & 5.12E+09 & 0.0435 \\
2MASX J12294019-4007220  & -111. & 0.468 & -17.20 & 4.45E+08 & 0.0012 & -143. & 0.446 & -17.22 & 4.54E+08 & 0.0435 \\
ESO 322- G 007           & -26. & 0.606 & -18.79 & 1.92E+09 & 0.0050 & -58. & 0.598 & -18.81 & 1.96E+09 & 0.0435 \\
MCG -06-27-023           & 176. & 0.563 & -19.91 & 5.40E+09 & 0.0140 & 144. & 0.589 & -19.93 & 5.51E+09 & 0.0435 \\
ESO 321- G 021           & -34. & 0.752 & -21.34 & 2.02E+10 & 0.0524 & -66. & 0.768 & -21.36 & 2.06E+10 & 0.0435 \\
ESO 322- G 019           & -123. & 0.660 & -19.81 & 4.92E+09 & 0.0128 & -155. & 0.642 & -19.83 & 5.02E+09 & 0.0435 \\
ESO 321- G?026           & -108. & 0.658 & -19.26 & 2.97E+09 & 0.0077 & -140. & 0.690 & -19.28 & 3.03E+09 & 0.0435 \\
ESO 322- G 020           & 200. & 0.674 & -20.29 & 7.66E+09 & 0.0199 & 168. & 0.656 & -20.31 & 7.81E+09 & 0.0435 \\
2MASX J12185570-4005358  & -28. & 0.810 & -- & -- & 0.0000 & -60. & 0.832 & -- & -- & 0.0435 \\
NGC 4499                 & 506. & 0.671 & -21.21 & 1.79E+10 & 0.0465 & 474. & 0.652 & -21.23 & 1.82E+10 & 0.0435 \\
2MASX J12301164-3845537  & 260. & 0.659 & -- & -- & 0.0000 & 228. & 0.664 & -- & -- & 0.0435 \\
ESO 322- G 017           & 175. & 0.692 & -19.03 & 2.40E+09 & 0.0062 & 143. & 0.706 & -19.05 & 2.45E+09 & 0.0435 \\
ESO 322- G 024           & -66. & 0.715 & -18.21 & 1.13E+09 & 0.0029 & -98. & 0.701 & -18.23 & 1.15E+09 & 0.0435 \\
 & & & & & & & & & &\\
RR242\_08064 & -79. & 0.222 & -15.75 & 1.17E+08 & 0.0005 & 85. & 0.196 & -15.65 & 1.07E+08 & 0.0357 \\
RR242\_13326 & -348. & 0.095 & -17.45 & 5.61E+08 & 0.0026 & -184. & 0.075 & -17.35 & 5.10E+08 & 0.0357 \\
RR242\_15689 & 211. & 0.108 & -15.87 & 1.31E+08 & 0.0006 & 375. & 0.096 & -15.77 & 1.19E+08 & 0.0357 \\
RR242\_20075 & -475. & 0.012 & -17.27 & 4.75E+08 & 0.0022 & -311. & 0.043 & -17.17 & 4.32E+08 & 0.0357 \\
RR242\_22327 & -283. & 0.068 & -18.54 & 1.53E+09 & 0.0070 & -119. & 0.084 & -18.44 & 1.39E+09 & 0.0357 \\
RR242\_23187 & -120. & 0.135 & -16.23 & 1.82E+08 & 0.0008 & 44. & 0.126 & -16.13 & 1.66E+08 & 0.0357 \\
RR242\_24352 & -823. & 0.106 & -19.89 & 5.31E+09 & 0.0243 & -659. & 0.118 & -19.79 & 4.82E+09 & 0.0357 \\
RR242\_25575 & -865. & 0.080 & -17.18 & 4.37E+08 & 0.0020 & -701. & 0.104 & -17.08 & 3.98E+08 & 0.0357 \\
RR242\_28727 & -503. & 0.115 & -16.55 & 2.45E+08 & 0.0011 & -339. & 0.118 & -16.45 & 2.23E+08 & 0.0357 \\
RR242\_36267 & -523. & 0.138 & -14.93 & 5.51E+07 & 0.0003 & -359. & 0.169 & -14.83 & 5.01E+07 & 0.0357 \\
NGC~5090 (RR~242a)      & -99. & 0.096 & -22.76 & 7.46E+10 & 0.3413 & 65. & 0.098 & -22.66 & 6.78E+10 & 0.0357 \\
NGC~5091 (RR~242b)      &   9. & 0.105 & -21.17 & 1.73E+10 & 0.0789 & 173. & 0.109 & -21.07 & 1.57E+10 & 0.0357 \\
NGC~5082                & 376. & 0.036 & -21.55 & 2.45E+10 & 0.1120 & 540. & 0.049 & -21.45 & 2.23E+10 & 0.0357 \\
NGC~5090B               & 748. & 0.100 & -20.89 & 1.33E+10 & 0.0610 & 912. & 0.133 & -20.79 & 1.21E+10 & 0.0357 \\
2MASX~J13201668-4327195 & -453. & 0.150 & -19.58 & 3.99E+09 & 0.0182 & -289. & 0.104 & -19.48 & 3.63E+09 & 0.0357 \\
NGC 5090A                & -93. & 0.114 & -21.34 & 2.02E+10 & 0.0923 & 71. & 0.106 & -21.24 & 1.83E+10 & 0.0357 \\
ESO 270- G 007           & 230. & 0.336 & -20.63 & 1.05E+10 & 0.0480 & 394. & 0.308 & -20.53 & 9.54E+09 & 0.0357 \\
2MASX J13181305-4330182  &   6. & 0.263 & -18.50 & 1.48E+09 & 0.0067 & 170. & 0.237 & -18.40 & 1.34E+09 & 0.0357 \\
AM 1317-425              & -197. & 0.371 & -- & -- & 0.0000 & -33. & 0.314 & -3.50 & 1.47E+03 & 0.0357 \\
2MASX J13195606-4306498  & -156. & 0.361 & -- & -- & 0.0000 & 8. & 0.305 & -3.50 & 1.47E+03 & 0.0357 \\
2MASX J13192359-4417358  & -162. & 0.376 & -18.70 & 1.77E+09 & 0.0081 & 2. & 0.394 & -18.60 & 1.61E+09 & 0.0357 \\
2MASX J13212941-4248564  & -57. & 0.555 & -17.60 & 6.44E+08 & 0.0029 & 107. & 0.489 & -17.50 & 5.85E+08 & 0.0357 \\
2MASX J13161705-4307595  & 390. & 0.566 & -18.80 & 1.94E+09 & 0.0089 & 554. & 0.516 & -18.70 & 1.77E+09 & 0.0357 \\
ESO 269- G 076           & -537. & 0.560 & -19.10 & 2.56E+09 & 0.0117 & -373. & 0.539 & -19.00 & 2.33E+09 & 0.0357 \\
2MASX J13270026-4359472  & -342. & 0.756 & -- & -- & 0.0000 & -178. & 0.726 & -3.50 & 1.47E+03 & 0.0357 \\
ESO 270- G 014           &  328. & 0.939 & -20.73 & 1.15E+10 & 0.0526 & 492. & 0.903 & -20.63 & 1.05E+10 & 0.0357 \\
ESO 269- G 072           & -220. & 0.755 & -21.26 & 1.87E+10 & 0.0857 & -56. & 0.714 & -21.16 & 1.70E+10 & 0.0357 \\
ESO 269- G 069           & -555. & 0.781 & -20.13 & 6.62E+09 & 0.0303 & -391. & 0.741 & -20.03 & 6.02E+09 & 0.0357 \\
\hline
\end{tabular}
\end{scriptsize}
\end{table*}

\end{appendix}

\end{document}